\documentclass[aps,pre,twocolumn,groupedaddress,showpacs,floatfix]{revtex4}
\usepackage{amsfonts}
\usepackage{graphicx}
\usepackage{color}
\usepackage{hyperref}

\newcommand{\vdimer}{{\vrule height0.2cm width0.05cm depth0pt}}
\newcommand{\hdimer}{{\hrule height0.05cm width0.2cm depth0pt}}

\newcommand{\hdimers}{{\hrule height0.05cm width0.1cm depth0pt}}
\newcommand{\verdimers}{\hbox{\vdimer \hskip 0.1cm \vdimer}}
\newcommand{\hordimers}{\hbox{\vbox{\hdimer \vskip 0.1cm \hdimer}}}
\newcommand{\vdimera}{\hbox{\vdimer \hskip 0.01cm}}
\newcommand{\hdimera}{\hbox{\vbox{\hdimer \vskip 0.05cm }}}

\newcommand{\hdimerb}{\hbox{\vbox{\hdimers \vskip 0.05cm }}}

\begin{document}
\title{Classical dimers with aligning interactions on the square lattice}

\author{Fabien Alet$^{1,2}$}
\email{alet@irsamc.ups-tlse.fr}
\author{Yacine Ikhlef$^{3,2}$}
\author{Jesper Lykke Jacobsen$^{3,2}$}
\author{Gr\'egoire Misguich$^2$}
\author{Vincent Pasquier$^2$}

\affiliation{$^1$Laboratoire de Physique Th\'eorique, UMR CNRS 5152,
  Universit\'e Paul Sabatier, 31062 Toulouse, France}
\affiliation{$^2$Service de Physique Th\'eorique, URA CNRS 2306, CEA Saclay, 91191 Gif sur Yvette, France}
\affiliation{$^3$LPTMS, UMR CNRS 8626, Universit\'e Paris-Sud, 91405 Orsay, France}

\date{\today}

\begin{abstract}

We present a detailed study of a model of close-packed dimers on the
square lattice with an interaction between nearest-neighbor dimers.
The interaction favors parallel alignment of dimers, resulting in a
low-temperature crystalline phase. With large-scale Monte Carlo and Transfer
Matrix calculations, we show that the crystal melts through a
Kosterlitz-Thouless phase transition to give rise to a high-temperature
critical phase, with algebraic decays of correlations functions with exponents
that vary continuously with the temperature. We give a theoretical
interpretation of these results by mapping the model to a Coulomb gas, whose
coupling constant and associated exponents are calculated numerically with
high precision.  Introducing monomers is a marginal perturbation at the Kosterlitz-Thouless transition
and gives rise to another critical line. We study this line numerically, showing that it
is in the Ashkin-Teller universality class, and terminates in a tricritical point at finite
temperature and monomer fugacity. In the course of this work, we also derive analytic results
relevant to the non-interacting case of dimer coverings, including a Bethe
Ansatz (at the free fermion point) analysis, a detailed discussion of the
effective height model and a free field analysis of height fluctuations.
\end{abstract}

\pacs{05.20.-y, 05.50.+q, 64.60.Cn, 64.60.Fr, 64.60.Kw}

\maketitle

\section{Introduction}
\label{sec:intro}

The problem of lattice coverings by "hard" objects, and dimers in particular,
is ubiquitous in classical statistical mechanics.
The formulation of the problem of dimer coverings goes back to the
thirties~\cite{Roberts}. The combinatorial problem of finding the exact
number of such coverings has been solved in the early sixties for planar 2d
lattices by means of Pfaffian techniques~\cite{Kasteleyn,Fisher}, which have
been extended to calculate dimer-dimer and monomer (i.e. unpaired
sites)-monomer correlation functions~\cite{FisherStephenson}. Notwithstanding
the intrinsic mathematical beauty of this problem~\cite{maths}, it also plays
a central role in statistical physics due to its relationship to
Ising~\cite{Fisher} or height models~\cite{Blote}. Dimer coverings of
bipartite graphs in three dimensions have also been recently shown to be
connected to gauge theories~\cite{Huse}. Dimer models have also recently regained interest because Quantum Dimer
Models (QDM), originally introduced by Rokhsar and Kivelson~\cite{rk}, are among the simplest systems
which exhibit ground-states with topological order and fractionalization~\cite{topo}.

\begin{center}
\begin{figure}
\includegraphics[height=6cm]{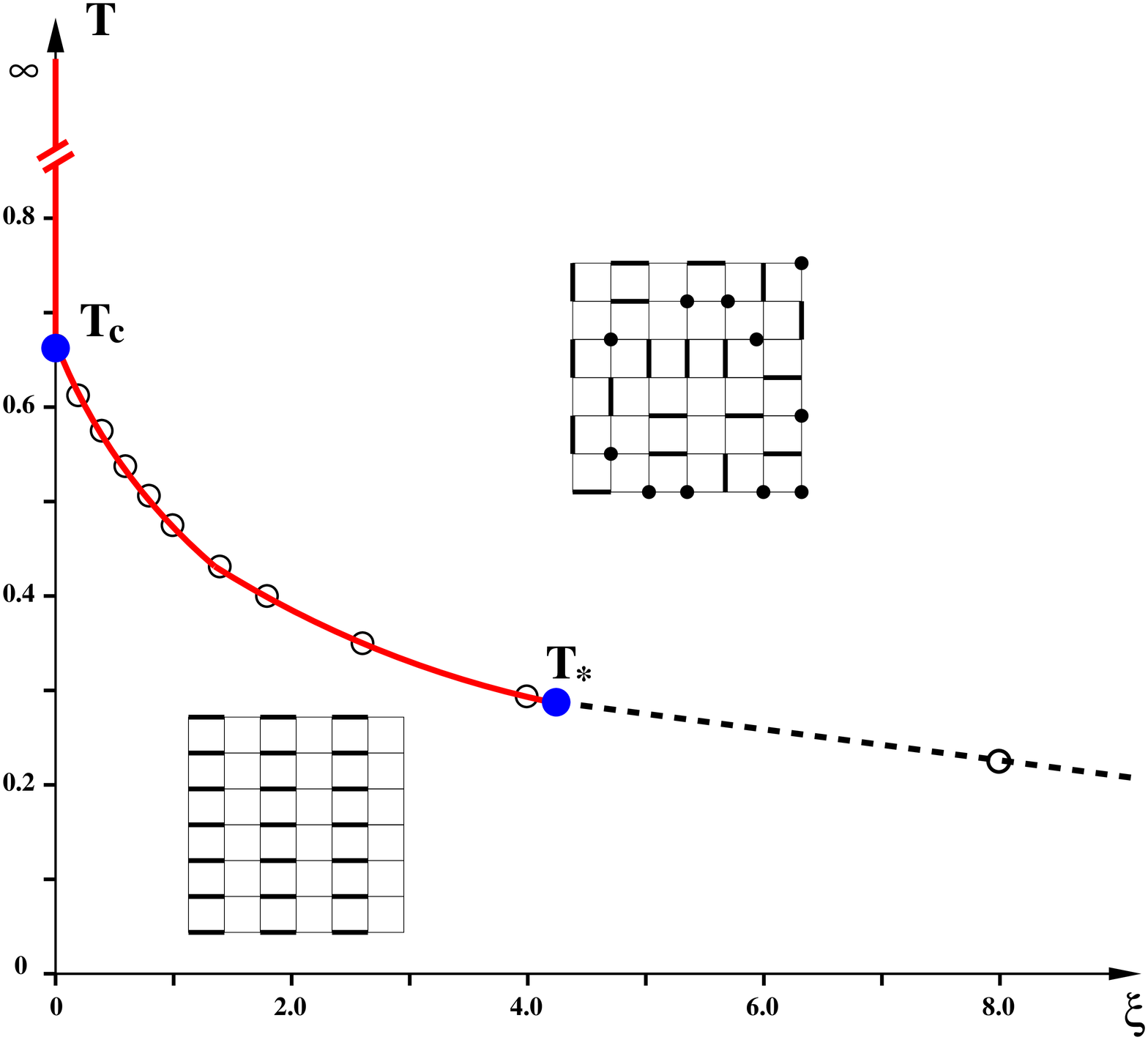}
\caption{(Color online) Phase diagram of the interacting dimer model in the
  temperature $T$, monomer fugacity $\xi$ plane (see text for definitions). The solid
  lines represent second-order phase transition lines with continuously
  varying exponents. When no monomers are allowed ($\xi=0$), the first
  critical line terminates at $T_c=0.65(1)$ and
  separates the high-$T$ critical phase from a long-range order crystalline
  phase through a Kosterlitz-Thouless phase transition. Allowing for
  monomers ($\xi \neq 0$) creates the second critical line separating the
  low $T$ crystalline phase from a monomer-dimer (massive) liquid phase at
  high $T$. This line terminates in a multicritical point at
  $T_\star=0.29(2)$, where it changes nature to become a first order
  line (dashed line). Simple energetic arguments (see Ref.~\onlinecite{short}) predict that the first order transition temperature scales as $1/(2\ln(\xi))$ when $\xi\rightarrow \infty$.
   }
\label{fig:pd}
\end{figure}
\end{center}

In this work, we study a model of {\it interacting} classical dimers on
the square lattice, with an interaction that favors dimer alignment. The dimer
coverings are close-packed, {\it i.e.} there are no sites left uncovered by a
dimer (monomers). We now describe the plan of the paper: we first introduce
the model and its simple limits in
Sec.~\ref{sec:model}. In Sec.~\ref{sec:bethe} we introduce the transfer matrix
of the model and describe how its critical exponents in the non-interacting
(infinite temperature) limit can be rederived by the Bethe Ansatz technique.
Unfortunately, the interacting model does not seem to be integrable by a
straightforward extension of this approach. We therefore go on, in
Sec.~\ref{sec:num}, to describe two complementary numerical simulation
schemes: Monte Carlo (MC) and Transfer Matrix (TM) calculations. The
results of the MC simulations are presented in Secs.~\ref{sec:columnar},
\ref{sec:KT} and \ref{sec:highT}: we find that the model possesses a
low-temperature crystalline phase separated by a Kosterlitz-Thouless (KT)
transition~\cite{KT} from a high-temperature critical phase with floating
exponents. We account for all these findings in Sec.~\ref{sec:theo}, where we
give a theoretical interpretation in terms of a Coulomb gas (CG)
picture~\cite{CG}. This mapping moreover allows use to make specific
predictions on the high-temperature phase that are successfully tested with
high-precision TM and MC calculations. The CG description implies that the
introduction of monomers is a marginal perturbation at the KT point and hence leads to the emergence
of another critical line. We study this numerically and find it to be in the
Ashkin-Teller universality class; the line terminates in a tricritical point
at finite temperature and monomer fugacity. The finally obtained phase diagram
is presented in Fig.~\ref{fig:pd}. We finally discuss the connections to
other models in classical statistical physics and the implications of our
findings for quantum models in Sec.~\ref{sec:discuss}, and conclude in
Sec.~\ref{sec:conc}. A short account of the results presented here was given
in Ref.~\onlinecite{short}.

\section{The model}
\label{sec:model}

We study a model of interacting close-packed dimers on the square lattice,
defined in the following way
\begin{eqnarray}
\label{eq:model}
Z & = & \sum_{c} \exp(- E_c/T) \nonumber \\
E_c & = & v ( ( N^c(\hordimers ) +  N^c(\verdimers ) ).
\end{eqnarray}
The sum in the partition function $Z$ is over all fully-packed dimer coverings
of the square lattice ${c}$. To each dimer covering $c$, we assign the energy
$E_c$ which simply counts the number $N^c(\hordimers)+ N^c(\verdimers)$ of
plaquettes with parallel (horizontal or vertical) dimers in the covering $c$.
$|v|=1$ sets the energy scale ($T$ is the temperature). The sign of $v$ determines the nature of the
interactions between the nearest-neighbor dimers : $v<0$ correspond to {\it
aligning} interactions between dimers, $v>0$ favors configurations
with staggered occupation of dimers. In this paper, we will consider $v=-1$,
the so-called columnar case. 

\begin{center}
\begin{figure}
\includegraphics[height=3cm]{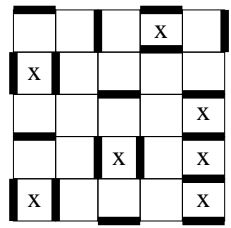}
\caption{Illustration of the interacting dimer model :  we consider dimer
coverings of the square lattice where each plaquette (marked with a cross)
with a pair of parallel nearest neighbor dimers contributes $+v$ to the energy
(the energy of this dimer covering is $7v$).}
\label{fig:model}
\end{figure}
\end{center}

\begin{center}
\begin{figure}
\includegraphics[width=7.6cm]{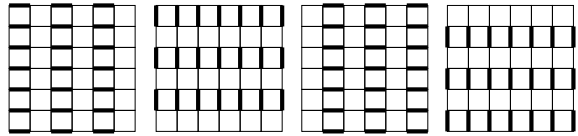}
\caption{The four columnar ground states.}
\label{fig:GS}
\end{figure}
\end{center}

The   model is illustrated     in Fig.~\ref{fig:model}, where a  dimer
covering  of   the  lattice   is   represented, and  the    plaquettes
contributing a  factor $+v$ to the   energy of this  configuration are
identified  by a  cross.   It is  straightforward to see   that at zero
temperature $T=0$, the configurations that minimize the energy are the
four states  represented  in Fig.~\ref{fig:GS}, where  the  dimers are
aligned in  {\it  columns}.  This  four-fold  degenerate ground  state
spontaneously breaks   translation and  $\pi/2$-rotational symmetries.
The   first  excitation above  these   ground-states are   obtained by
flipping two parallel dimers around a plaquette;  the system has a gap
(it costs a finite  energy  $2v$  to flip   the  two dimers) and   the
columnar order is therefore expected to  subsist at (possibly  small
but) finite temperature.

On the other hand, the partition function at infinite temperature $T=\infty$
is simply the unweighted sum over all possible dimer coverings of the square
lattice, and the model can be solved exactly at this
point~\cite{Kasteleyn,Fisher}. The $T=\infty$ point is critical, with
correlation functions displaying an algebraic dependence with
distance~\cite{FisherStephenson}: dimer-dimer correlation functions decay as
$1/r^2$ and monomer-monomer correlation functions as $1/\sqrt{r}$ for large
distance $r$. We postpone the precise definitions of these correlation
functions to Sec.~\ref{sec:bethe} below where we rederive the results for the
critical exponents using another exact approach, the coordinate Bethe
Ansatz. Although we have not been able to solve the interacting dimer problem (finite temperature),
the Bethe Ansatz technique can potentially go beyond free fermion problems
(contrary to the  Pfaffian  methods    of
Refs.~\onlinecite{Kasteleyn,Fisher,FisherStephenson}).

The Bethe Ansatz method also serves to illustrate that the critical nature of the dimer
covering problem is intimately linked to the bipartite nature of the square
lattice (non-bipartite lattices present a dimer liquid behavior with a finite
correlation length~\cite{topo}). Unfortunately, the introduction of interactions appears
to break the integrability of the model.

We    end up  the  introduction  with   some historical notes  on this
model. The model Eq.~(\ref{eq:model})    was first introduced in   the
physics of  liquid  crystals~\cite{old} and  not  developed further in
this context  to our best  knowledge. This is likely  due to  the fact
that the quest was  there to look for microscopic  models where a true
liquid crystal phase exists,  and not a crystalline  state such as the
one     depicted  in   Fig.~\ref{fig:GS}.   Later   on,  Brankov   and
coworkers~\cite{Brankov} also studied the  same model with Monte Carlo
methods  but missed  the true  critical  behavior of this problem.
In a recent  publication~\cite{short} we described  the physics of the
undoped  model and  sketched the  existence  of a  critical line  with
central charge $c=1$ ending at a tri-critical point at finite doping.
This has then been followed  by  other studies, including
construction  of   quantum   models~\cite{Castelnovo,Poilblanc}   with
ground-state  wave-functions  described by   the partition function in
Eq.~(\ref{eq:model}), further investigations of
the  doped monomer case~\cite{Castelnovo,Poilblanc,Papa} and generalization to three-dimensional
lattices~\cite{Alet3d}.

\section{Non-interacting dimers as free fermions}
\label{sec:bethe}

We study dimer coverings (each site is paired  with exactly one of its
neighbors) of the square lattice. In this section (and only in this section), we give a fugacity $\omega$ to each
horizontal dimer, and $1$ to  vertical ones.
This is the model of \emph{non-interacting} dimers,
solved by combinatorial methods \cite{Kasteleyn, Fisher,
FisherStephenson}. We will introduce the TM of this
model and we show how to compute the partition sum and
the correlation functions by the Bethe Ansatz method.

\subsection{Transfer matrix}
\label{sec:bethe:tm}

\begin{figure}
\begin{center}
\includegraphics{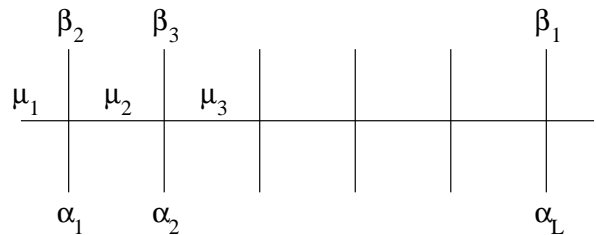}
\end{center}
\caption{The row-to-row transfer matrix}
\label{fig:transfer}
\end{figure}

The partition sum of the model is:
\begin{equation}
Z = \sum_{\mathrm{dimer\ config.}} \omega^{\# \mathrm{horizontal\ dimers}}
\end{equation}
On a strip of width $L$, we define the state of a row as the ``occupation
numbers'' $\alpha=(\alpha_1, \dots, \alpha_L)$ of the vertical edges, where
$\alpha_i$ is equal to $1$ if the $i$-th vertical edge is occupied by a dimer,
and $0$ otherwise. There are $2^L$ configurations of a row, so $Z$ can be
written as the trace of a $2^L$-dimensional transfer matrix $T$. We impose
periodic boundary conditions (PBC), {\it i.e.} the index $i$ is considered modulo $L$.
Given two line configurations $\alpha$ and $\beta$, the matrix element
$T_{\beta \alpha}$ is the sum of the Boltzmann weights associated with the
horizontal dimer configurations $\mu$ compatible with $\alpha$ and
$\beta$:
\begin{equation}
T_{\beta \alpha} = 
  \sum_{\mu | (\alpha ,\beta)} \omega^{\mu_1 + \dots + \mu_L}
\end{equation}
as illustrated in Fig.~\ref{fig:transfer}. The compatibility criterion $\mu |
(\alpha ,\beta)$ can be expressed formally as follows:
\begin{equation}
 \forall i \in \{1,\ldots,L\}~: \quad
 \mu_i + \mu_{i+1} + \alpha_i + \beta_{i+1} = 1.
\end{equation}

\subsection{Conservation law} 
\label{sec:bethe:conservation}

\begin{figure}
\begin{center}
\includegraphics[scale=0.4]{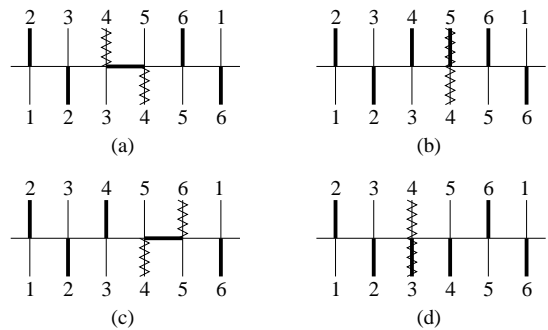}
\end{center}
\caption{Rules for the dynamics of the particles on a strip of 
width $L=6$. Particles are represented by zigzag lines. The rules 
are different for particles starting from an even site (a, b, c) and 
an odd site (d).}
\label{fig:particle}
\end{figure}

For convenience, we introduce a shift at each row in the numbering 
of columns (see Fig.~\ref{fig:transfer}). We call \emph{particle} 
an empty even vertical edge \emph{or} an occupied odd vertical edge.
Let us show that the number of particles is conserved, and let us give 
at the same time the rules for the dynamics of the particles (see the
corresponding Fig.~\ref{fig:particle}).

\textit{Particle on an even vertical edge:} If a particle sits on the
vertical edge $\alpha_{2j}$, then $\alpha_{2j}=0$. The site above this
edge must be visited once, so one of the variables $\mu_{2j}$,
$\beta_{2j+1}$, $\mu_{2j+1}$ must be equal to 1. In the first 
(resp. third) case, this implies that $\beta_{2j}$ (resp. $\beta_{2j+2}$)
is zero. Therefore, in the next row, there is a particle 
on the edge $\beta_{2j}$, $\beta_{2j+1}$ or
$\beta_{2j+2}$.

\textit{Particle on an odd vertical edge:} If a particle sits on the
vertical edge $\alpha_{2j-1}$, then $\alpha_{2j-1}=1$. The site above 
this edge has already been visited, so $\beta_{2j}$ must be zero:
in the next row, the particle sits on the edge $\beta_{2j}$.

The TM is block-diagonal, each block representing
a sector with fixed number of particles $n$. We call $T^{(n)}$
the TM block in the $n$-particle sector. Note that the lattice width
$L$ must be even, because, for an odd lattice width with PBC, the
number of particles is not conserved.

\subsection{One-particle sector}
\label{sec:oneparticle}

The action of $T$ on a one-particle state $\Phi$ is:
\begin{equation}
\begin{array}{lccl}
(T \Phi) & (2j) & = & \omega \Phi(2j-2) + \Phi(2j-1) + \omega \Phi(2j) \\
(T \Phi) & (2j+1) & = & \Phi(2j)
\end{array}
\end{equation}
We want to take advantage of the translational invariance to diagonalize
$T^{(1)}$. Define the two-step cyclic permutation $J$ of the sites by
its action on a one-particle state $\Phi$: 
\begin{equation}
(J \Phi)(x) = \Phi(x+2)
\end{equation}
On a lattice of even width with PBC, the operator
$J$ commutes with $T^{(1)}$. If $z$ satisfies the condition $z^L=1$,
the eigenspace of $J$ with eigenvalue $z^2$ is generated by the 
two vectors $\Phi_z, \overline{\Phi}_z$:
\begin{equation}
\begin{array}{lcl}
\Phi_z(2j)= z^{2j} & , & \Phi_z(2j-1)= 0 \\
\overline{\Phi}_z(2j)=0 & , & \overline{\Phi}_z(2j-1)= z^{2j-1}
\end{array}
\end{equation}
Note that $\Phi_{-z} = \Phi_z$ and
$\overline{\Phi}_{-z} = -\overline{\Phi}_z$.

More generally, let $z$ be a complex number of modulus unity.
The block of $T^{(1)}$ in the basis 
$(\Phi_z, \overline{\Phi}_z)$ is:
\begin{equation}
\left[
\begin{array}{cc}
\omega \ (1+z^{-2}) & z^{-1} \\
z^{-1}              & 0
\end{array}
\right]
\end{equation}
This matrix has eigenvectors $\psi_z, \psi'_z$ with the
eigenvalues $\Lambda(z), \Lambda'(z)$ satisfying:
\begin{equation} \label{eq:lambda}
\begin{array}{ccl}
\Lambda(z) + \Lambda'(z) &=& \omega \ (1+z^{-2}) \\
\Lambda(z) \ \Lambda'(z) &=& -z^{-2}
\end{array}
\end{equation}
One can then write:
\begin{eqnarray}
z &=& \exp \, (ik) \\
\Lambda(z)  &=& \exp \, [h+i(\phi+\theta)] \\
\Lambda'(z) &=& \exp \, [-h+i(\phi-\theta)] 
\end{eqnarray}
where $k, \phi, \theta$ are real and $h$ is nonnegative. $\Lambda(z)$ is
the eigenvalue with greatest modulus.
With this parametrization, Eqs.~(\ref{eq:lambda}) imply:
\begin{equation}
\cosh(2h)= 1 + 2 \omega^2 \cos^2 k
\end{equation}
when $|z|=1$. Recalling that $\cos k \geq 0$, this relation can 
be inverted and we obtain:
\begin{equation} \label{eq:h}
h(k) = 
\log \left[ \omega \cos k + (1+ \omega^2 \cos^2 k)^{\frac{1}{2}} \right]
\end{equation}
when $-\pi/2 \leq k  \leq \pi/2$. A useful quantity for the computation of 
finite-size effects is $h'(\pi/2) = -\omega$.

\subsection{Two-particle sector, scattering amplitude}

Consider the action of the TM on the two-particle vector:
\begin{equation}
\psi_{12}(x_1, x_2) = \psi_{z_1}(x_1) \psi_{z_2}(x_2) \ , \qquad x_1<x_2
\end{equation}
The TM changes the positions of the particles from $(x_1, x_2)$
to $(y_1, y_2)$. For fixed positions $y_1< y_2$, let us look at the
initial states leading to these positions:
\begin{itemize}
\item if $y_2 > y_1 + 2$ or $(y_1,y_2)=(2j-1,2j+1)$, then for each
particle all initial states are allowed, thus:
\begin{eqnarray*}
(T \psi_{12})(y_1, y_2) & = &
\sum_{x_1,x_2} \psi_{z_1}(x_1) \psi_{z_2}(x_2) T_{y_1, x_1} T_{y_2, x_2} \\
\ & = & \Lambda(z_1) \Lambda(z_2) \psi_{z_1}(y_1) \psi_{z_2}(y_2) 
\end{eqnarray*}
\item if $(y_1, y_2)=(2j-1,2j)$, all initial states 
are allowed except $x_1=x_2=2j-2$. One has to subtract
the corresponding term in the action of the matrix $T$:
\begin{eqnarray*}
(T \psi_{12})(2j-1, 2j) & = & 
\Lambda(z_1) \Lambda(z_2) \psi_{z_1}(2j-1) \psi_{z_2}(2j) \\
     & \ & - \omega \psi_{z_1}(2j-2) \psi_{z_2}(2j-2)
\end{eqnarray*}
\item if $(y_1, y_2)=(2j,2j+1)$ or $(y_1, y_2)=(2j,2j+2)$, all 
initial states are allowed except $x_1=x_2=2j$. Similarly to the previous case:
\begin{eqnarray*}
(T \psi_{12})(2j, 2j+1) & = &
\Lambda(z_1) \Lambda(z_2) \psi_{z_1}(2j) \psi_{z_2}(2j+1) \\
     & \ & - \omega \psi_{z_1}(2j) \psi_{z_2}(2j)
\end{eqnarray*}
\begin{eqnarray*}
(T \psi_{12})(2j, 2j+2) & = &
\Lambda(z_1) \Lambda(z_2) \psi_{z_1}(2j) \psi_{z_2}(2j+2) \\
    & \ & - \omega \psi_{z_1}(2j) \psi_{z_2}(2j)
\end{eqnarray*}
\end{itemize}
Note that all the interaction terms in $T \psi_{12}$ are symmetric functions
of the momenta $k_1, k_2$. As a consequence, the antisymmetric combination:
\begin{equation}
\psi(x_1,x_2) = \psi_{z_1}(x_1) \psi_{z_2}(x_2) - \psi_{z_2}(x_1) \psi_{z_1}(x_2)
\end{equation}
is an eigenvector of the matrix $T$ with eigenvalue $\Lambda(z_1) \Lambda(z_2)$.

\subsection{Periodic boundary conditions, position of the solutions}
\label{sec:bethe.sol}

\begin{figure}
\begin{center}
\includegraphics[scale=0.5]{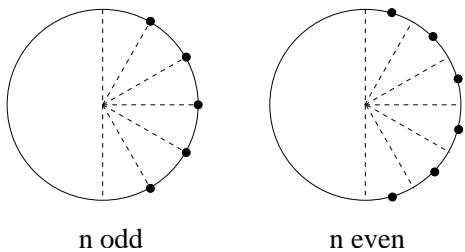}
\end{center}
\caption{Position of the vacancies for a strip of width $L=12$.
Vacancies are represented by black points on the unit circle, in
the complex $z$ plane.}
\label{fig:circle}
\end{figure}

The analogous construction in the $n$-particles sector
gives the eigenvectors and eigenvalues:
\begin{eqnarray}
\psi(x_1, \dots, x_n) & = &
  \sum_P \epsilon(P) \psi_{z_{p_1}}(x_1) \dots \psi_{z_{p_n}}(x_n) \\
\Lambda(z_1, \dots z_n) & = & \Lambda(z_1) \dots \Lambda(z_n)
\end{eqnarray}
where the sum is over all permutations of the integers $1, \dots, n$ and
$\epsilon(P)$ is the signature of the permutation $P$. PBC yield:
\begin{equation}
\forall j \quad (z_j)^L = (-1)^{n-1}
\end{equation}
The solutions of these equations lie on the unit circle, which justifies
the discussion in section~\ref{sec:oneparticle}. The momenta $k_j$ are
given by:
\begin{equation}
\begin{array}{rlll}
k_j = & \frac{2 \pi}{L} I_j, & \ I_j \in \mathbb{Z} & n \ \mathrm{odd} \vspace{2mm} \\
k_j = & \frac{2 \pi}{L} \left( I_j+\frac{1}{2} \ \right), & 
      \ I_j \in \mathbb{Z} & n \ \mathrm{even}
\end{array}
\end{equation}
See Fig.~\ref{fig:circle} for a graphical representation of the
vacancies on the unit circle.
For any $z$ on the unit circle, the TM has 
an eigenvalue $\Lambda$ with modulus greater than one. Therefore, 
the number of particles that maximizes the total eigenvalue of $T$ 
is either the greatest even value or the greatest odd value for $n$.
Since the $k_j$'s are distinct, lie in the interval 
$[-\pi/2, \pi/2]$ and are spaced by $2 \pi/L$, the maximum number
of particles is $L/2$.

\subsection{Thermodynamic limit}
\label{sec:bethe.TL}

In this section, the discussion is restricted for simplicity to a system of width 
multiple of four: $L=4p$. According to the previous section, the
leading sector is defined by the greatest eigenvalue either in 
the sector $n=2p$ or in the sector $n=2p-1$. As will be shown in a few lines, the correct choice for the leading 
sector is $n=2p$.

For a system of finite width and infinite length,
the free energy density per surface unit in the $n$-particles sector is 
defined by:
\begin{equation}
f_L^{(n)} = \frac{1}{L} \log \Lambda_{\mathrm{max}}^{(n)}
\end{equation}
where $\Lambda_{\mathrm{max}}^{(n)}$ is the eigenvalue of $T$ with greatest
modulus in the $n$-particles sector. The corresponding quantity for
$L=4p$ and $n=2p$ is:
\begin{equation}
f_L^{(L/2)} = \frac{2}{L} \sum_{j=0}^{p-1} h \left[ (j+1/2) \frac{2 \pi}{L}\right]
\end{equation}
where the function $h(k)$ is given by Eq.~(\ref{eq:h}).
When $L$ goes to infinity, this quantity tends to the limit $f_\infty$: 
\begin{equation}
f_\infty = \frac{1}{\pi} \int_0^{\frac{\pi}{2}} 
\log \left[ \omega \cos k + (1+ \omega^2 \cos^2 k)^{\frac{1}{2}} \right]
\ dk
\end{equation}
in agreement with formula (17) of Ref.~\onlinecite{Kasteleyn}.
In the isotropic case $\omega=1$:
\begin{equation}
f_\infty(\omega=1) = \frac{2G}{\pi}
\end{equation}
where $G$ is the Catalan constant:
\begin{equation}
G = 1^{-2} - 3^{-2} + 5^{-2} - 7^{-2} + \dots
\end{equation}
The asymptotic behavior of $f_L^{(L/2)}$ is derived from the Euler-Maclaurin
formula:
\begin{equation}
f_L^{(L/2)} = f_\infty + \omega \frac{\pi}{6 L^2} + o \left( L^{-2} \right).
\end{equation}
We expect the critical point to have conformal symmetry in the isotropic
case, with a central charge $c=1$.
If one particle is removed ($n=2p-1$), the solutions $z_j$ all 
get shifted (see Fig.~\ref{fig:circle}). The Euler-Maclaurin formula yields:
\begin{equation}
f_{L}^{(L/2-1)} = f_\infty - \frac{\pi}{6L^2} h' \left( \frac{\pi}{2} \right)
-\frac{1}{\pi} I \left( \pi/L \right) + o \left( L^{-2} \right)
\end{equation}
with
\begin{equation}
I(\epsilon) = \int_{\frac{\pi}{2}-\epsilon}^{\frac{\pi}{2}} h(k) 
                  \ dk 
            = -h' \left( \frac{\pi}{2} \right) \epsilon^2 / 2 
              + o(\epsilon^2).
\end{equation}
Finally we obtain:
\begin{equation}
f_L^{(L/2-1)} = f_L^{(L/2)} -\frac{\pi}{2L^2} \omega + o \left( L^{-2} \right)
\end{equation}
This proves that in the thermodynamic limit the leading sector is indeed
given by $n=2p$. In the isotropic case, the critical exponent corresponding to
the removal of one particle is $X_1=1/4$ (see definition and discussion in Sec.~\ref{sec:TM} below).
Now if two particles are removed from the leading sector, the other $z_j$'s 
are not shifted, because the number of particles remains even.
The only effect is a decrease of free energy caused by the absence
of the two particles:
\begin{eqnarray}
f_L^{(L/2-2)} & = & f_L^{(L/2)} - \frac{2}{L} 
             h \left( \frac{\pi}{2} - \frac{\pi}{L} \right) \\
 \  & = & f_L^{(L/2)} - \frac{2 \pi}{L^2} \omega + o(L^{-2})
\end{eqnarray}
In the isotropic case, the critical exponent corresponding to this process (see again
Sec.~\ref{sec:TM} below) is $X_2=1$.

\subsection{Introducing interactions}
\label{sec:bethe:int_transfer}

\begin{figure}
\begin{center}
\includegraphics[width=6cm]{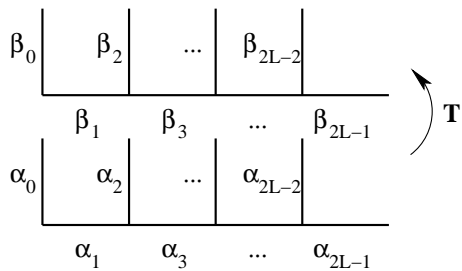}
\end{center}
\caption{The row-to-row transfer matrix for the interacting case.}
\label{fig:int_transfer}
\end{figure}

A TM for the interacting model Eq.~(\ref{eq:model}) can be written
down by generalizing the working of Sec.~\ref{sec:bethe:tm}. To this end, the
basis states $(\alpha)$ must encode not only the occupation numbers of a row
of vertical edges (as before), but also the occupation numbers of the
preceding row of horizontal edges, as shown in Fig.~\ref{fig:int_transfer}.

The transfer matrix $T$ is most easily defined by giving its sparse matrix
decomposition
\begin{equation}
 T = T_2 T_1, \qquad T_k = \prod_{i=0}^{L-1} T_k^{(i)},
 \label{sparse}
\end{equation}
where the matrix $T_1^{(i)}$ encodes the interactions at plaquette $i$
and $T_2^{(i)}$ imposes the dimer constraint at vertex $i$. More precisely,
$T_1^{(i)}$ evolves $\alpha_{2i+1}$ into $\beta_{2i+1}$ and has matrix
elements (we set $W \equiv {\rm e}^{-v/T}$) 
\[
 T_1^{(i)}(\alpha_{2i} \alpha_{2i+1} | \beta_{2i+1} \alpha_{2i+2}) =
 W (\alpha_{2i}\alpha_{2i+2} + \alpha_{2i+1}\beta_{2i+1}) \,,
\]
whereas $T_2^{(i)}$ evolves $\alpha_{2i}$ into $\beta_{2i}$ and has matrix
elements
\[
 T_2^{(i)}(\beta_{2i-1} \alpha_{2i} | \beta_{2i} \beta_{2i+1}) =
 \delta_{\beta_{2i-1} + \alpha_{2i} + \beta_{2i} + \beta_{2i+1},1} \,.
\]

We have attempted to diagonalize $T$ using the Bethe Ansatz method,
using a straightforward generalization of the working exposed in the
preceding subsections but we failed to obtain a
consistent determination of the scattering amplitudes $S(z_i,z_j)$.
Most likely, this means that the interacting dimer model is not integrable.
In the remainder of the paper, we therefore study the model using numerical
and (non-rigorous) field theoretical methods.

\section{Numerical methods}
\label{sec:num}

Our numerical methods consist in MC simulations and exact
diagonalization of the TM. We now describe these two methods in
turn.

\subsection{Monte Carlo calculations}
\label{sec:MC}

We use a MC directed-loop (or directed-``worm'') algorithm~\cite{Sandvik}. This method allows to make
non-local moves in the dimer configurations by changing the positions of
dimers along a closed loop, which can be quite large. This results in small
autocorrelation times in the MC process, and permits to treat large systems
(up to $512 \times 512$ in this study). Moreover, the directed-loop algorithm
captures the physics of test defects (monomers) in the dimer configuration as
we discuss below. The algorithm indeed allows to calculate
monomer-monomer correlation function; conversely, this indicates that the
performance of the algorithm is dictated by the physical properties of test
monomers in the different physical phases.

For sake of completeness, we briefly describe below the
algorithm following Ref.~\onlinecite{Sandvik} and specifying minor details where our specific
implementation differs. One MC sweep of the algorithm consists of the three following steps:

\begin{itemize}
\item[1.] The worm, which can be seen as constituted by two monomers (head and
tail), is initially placed on top of a dimer configuration at a random site
$i=i_0$.

\item[2.] The site $i$ is connected to a neighboring site $j$ by a dimer in
the background configuration (this dimer is noted $(i,j)$). The head of the worm is moved to $j$ and the
dimer ($i$,$j$) is removed, leaving the site $j$ with no dimer
attached to it. Out of the four neighbors of $j$, one (which we call $k$) is
selected according to a local detailed balance rule (see below). A dimer is
put between $j$ and $k$.

\item[3.] If $k=i_0$, the worm is finished and we are left with a new valid
dimer configuration. Otherwise, we rename $i=k$ and go back to step 2.

\end{itemize}

How does the worm, sitting at site $j$ (and coming from site $i$), choose the
site $k$ where a dimer will be put in step 2 ? For this, we consider the {\it
weights} $w_{(ij)}$ (respectively $w_{(jk)}$) contributed to the partition
function by a dimer located between sites $i$ and $j$ (respectively $j$ and
$k$). In the model of Ref.~\onlinecite{Sandvik}, each dimer is given a certain
fugacity and thus contributes solely a certain weight to the partition
function. In our model, the weight of a dimer $(i,j)$ is given by $w_{(ij)}=\exp(-v.N_{ij}/T)$ where $N_{ij}\in \{0,1,2\}$ is the number
of nearest neighbors parallel to the dimer $(i,j)$. Once these weights are known, the probability $P((i,j)\rightarrow (j,k))$ to
select a given site $k$ is imposed to satisfy a local detailed balance rule:

\begin{equation}
P((i,j)\rightarrow (j,k)) w_{(ij)} = P((j,k)\rightarrow (i,j)) w_{(jk)}.
\end{equation} 

This leads to a set of equations (``directed-loop'' equations~\cite{Sandvik})
corresponding to all the possibles values of local dimer configurations and
the corresponding numbers $N_{ij}$. These equations are underdetermined, and
we impose by experience~\cite{Sylju,Alet03} to minimize the bounce processes
$ P((i,j)\rightarrow (j,i))$, {\it i.e.} the case where the site $k$ is chosen to be
the origin site $i$ (the worm backtracks in its own path, which is {\it a
priori} quite useless). For the specific model of Ref.~\onlinecite{Sandvik}, a
solution minimizing the bounce probabilities and satisfying the local detailed
balance equation was found analytically. More generally, such a solution can
always be found numerically with linear programming techniques~\cite{Alet03}.
Please note that at $T=\infty$, the worm simply performs a random
walk in the dimer configuration (more precisely, all the even steps in the walk are purely random, the odd ones are dictated by the underlying dimer
configuration).

Taking a snapshot of the configuration during the worm construction shows that
two test monomers have been inserted in the dimer configuration, and thus
connect the behavior of the worm to the monomer correlation function. The fact
that the worm walk is {\it locally} detailed balance actually imposes the
histogram of the distance ${\bf r}$ between the worm's head and tail to be
proportional (up to a small correction factor) to the monomer-monomer correlation function $M({\bf r})$ (see
precise definition in Sec.~\ref{sec:highT}). The proof of this statement can
be worked out along the lines of Ref.~\onlinecite{Alet03}. The only
subtlety is the following: the measurement of the correlation function is
made at step 2, before the selection of the next site $k$. Since all
the future dimer positions $(j,k)$ are not equivalent (they will contribute
differently to the partition function) and since the next dimer position is not yet decided, we have to correct the
monomer-monomer correlation estimator by the inverse of
the total weight contributed by all possible future positions, {\it i.e.} we
increment the estimator of $M({\bf r})$  (with ${\bf r}$ the
position difference vector between sites $i_0$ and $j$) by $W_j^{-1}$ where $W_j=\sum_{k}w_{(jk)}$.
If all future position dimers are equivalent (as in the model of
Ref.~\onlinecite{Sandvik}), this factor is constant, and we can just simply
identify the histogram of the distance ${\bf r}$ between the worm's head and
tail to $M({\bf r})$.

The worm algorithm therefore possesses the nice feature of being able to calculate $M({\bf r})$, {\it even if monomers are not
allowed in the model}. This also indicates that the worm algorithm
performances is bound to follow the physics of monomers: if the
monomers are confined, the worms will be short, resulting in a merely local
algorithm - which is known to display poor performances (for example
ergodicity problems). If the monomers are deconfined, worms will be long and
will update a massive number of dimers - resulting in small autocorrelation
times.

The technical details of the MC calculations are as follows: simulations were
performed on $N=L \times L$ samples, up to $L=160$ for the full $T$ range,
and up to $L=512$ for correlation functions for a few chosen temperatures.
PBC are assumed. Each MC sweep is constituted by a
number of worms such that all the links of the lattice are visited once
on average by a worm. For each parameter set, a total between $10^6$ and $10^7$ sweeps
was performed.

\subsection{Transfer matrix calculations}
\label{sec:TM}

The TM for the interacting dimer model was defined above in
Sec.~\ref{sec:bethe:int_transfer}. We shall henceforth suppose that the width
$L$ of the lattice strip is even; the periodic boundary conditions in the
$L$-direction are then compatible with the bipartiteness of the lattice. By
virtue of the conservation law established in
Sec.~\ref{sec:bethe:conservation}, the TM has a block diagonal structure, with
each block corresponding to a fixed number of particles. It is convenient to
define a ``charge'' $Q$ corresponding to each block, as $Q=L/2-n$,
where $n$ is the number of particles. Also, we label the
eigenvalues $\Lambda^Q_k$ within each block in order of decreasing norm:
$|\Lambda^Q_1| \ge |\lambda^Q_2| \ge \ldots$.

\subsubsection{Correlation functions}

For entropic reasons, the largest eigenvalue must be located in the $Q=0$
block, $\Lambda_{\rm max} = \Lambda^0_1$. By the Perron-Frobenius theorem, it
corresponds to the unique eigenvector in which all entries are non-negative.
Consider first a dimer covering of a strip of size $L \times M$, with free
(resp.\ periodic) boundary conditions in the $M$ (resp.\ $L$) direction. Only
the TM eigenvalues of the $Q=0$ block will contribute to the corresponding
partition function $Z$. Let us now modify the problem by marking $Q_0 > 0$
vertices of the even sublattice in the bottom row, and $Q_0$ vertices of the
odd sublattice in the top row. In the modified problem, dimers are required to
cover all unmarked vertices and none of the marked vertices. The TM
eigenvalues contributing to the modified partition function $Z_{Q_0}$ are then
exactly those of the $Q=Q_0$ block. (For $Q_0 < 0$, interchange the two
sublattices and change the sign of $Q_0$).

Physically, the marked vertices can be interpreted as monomer defects in the
surrounding dimer environment. The ratios $C_{Q_0}(M) \equiv Z_{Q_0}/Z$ define
(unnormalized) correlation functions, measuring the correlations between the
two groups of monomers, separated by a distance $M$. For $M \gg L$ the
correlations decay exponentially as $C_{Q_0}(M) \sim
(\lambda^{Q_0}_1/\Lambda^0_1)^M$. 

If the system enjoys conformal invariance, this corresponds to an algebraic
decay in the plane. More precisely, define the free energies per unit area as
$f^Q_k = L^{-1} \log \Lambda^Q_k$. The finite-size dependence \cite{cardy_84}
\begin{equation}
 f^0_1 - f^{Q_0}_1 = \frac{2 \pi X_{Q_0}}{L^2} + o(L^{-2})
 \label{car84}
\end{equation}
then defines a critical exponent $X_{Q_0}$ whose interpretation reads as
follows: let ${\cal C}_N$ be a dimer covering of an $N \times N$ square with
free boundary conditions (planar geometry), with two small regions of
$Q_0$ monomer defects, each region corresponding to a definite sublattice as
above. Suppose that each region has an extent of the order of the lattice
spacing and is far from the boundaries. Then the probability that the two regions are separated by a distance
$r$ satisfying $1 \ll r \ll N$ is proportional to $r^{-2 X_{Q_0}}$.

The corresponding conformal field theory (CFT) is further characterized by
its central charge $c$, which is related to the finite-size dependence
of $\Lambda_{\rm max}$ as follows \cite{cardy_86}:
\begin{equation}
 f^0_1 = f_\infty + \frac{\pi c}{6 L^2} + o(L^{-2}) \,,
 \label{car86}
\end{equation}
where $f_\infty = \lim_{L\to\infty} f^0_1$ is the bulk free energy.

While $X_1$ determines the leading monomer-monomer correlation function,
the leading dimer-dimer correlation can be obtained from Eq.~(\ref{car84})
by replacing $f^{Q_0}_1$ by $f^0_2$.

\subsubsection{Numerical procedure}

The leading eigenvalue of a given block $Q$ is obtained by an
iterative procedure (the so-called power method \cite{Wilkinson}) in
which the relevant TM block $T_Q$ is multiplied onto a
vector of weights which is indexed by the basis states of that block.
This vector can be taken initially as a single arbitrary basis state,
which is known to belong to the block $Q$.  The eigenvalue
$\Lambda^Q_1$ is then related to the asymptotic growth of the norm of
the iterated vector.

This procedure has multiple practical advantages: (i) only the iterated
vector, and not $T_Q$ itself, needs to be stored in memory, (ii) using
the factorization Eq.~(\ref{sparse}) one can take advantage of sparse
matrix techniques, so that one iteration is performed in time $\sim L. {\rm dim}(T_Q)$, (iii) the complete state space corresponding to $T_Q$
is automatically generated in the iterative process. To store and
access the weights in an efficient manner ({\it i.e.} in constant time),
standard hashing techniques are employed.

To obtain higher eigenvalues, $\Lambda^Q_k$ with $k \ge 2$, one can
similarly iterate a set of vectors which is kept mutually orthogonal
at the end of each iteration \cite{Wilkinson}. Alternatively, one can
in some cases use the symmetry of the corresponding eigenvectors. As
an example of this, note that the eigenvectors corresponding to
$\Lambda^0_1$ (resp.~$\Lambda^0_2$) are even (resp.~odd) upon shifting
the lattice by one unit in the horizontal direction.

The computational effort needed to obtain the largest eigenvalue can be judged
from Table~\ref{tab:tmsize} which shows the size of the block $T_0$ for various strip widths $L$.
Note that these numbers increase much slower than the naive estimate
$4^L$, that one would obtain by considering the possible occupation
numbers while ignoring the dimer constraint and the value of $Q$.  We
limited the present study to $L_{\rm max}=18$, although a couple of
more sizes could have easily been obtained.

\begin{table}
 \begin{tabular}{l|rrrrrrrrr}
 $L$        & 2  &  4 &  6 &   8 &   10 &    12 &    14 &     16 &      18 \\
 \hline
 {\rm dim}($T_0$) 
            & 4  & 16 & 76 & 384 & 2004 & 10672 & 57628 & 314368 & 1728292 \\
 {\rm dim}($T_1$) 
            & 1  &  8 & 48 & 272 & 1520 &  8472 & 47264 & 264224 &  1480608\\
 {\rm dim}($T_2$) 
            &    &  1 & 12 &  96 &  660 &  4224 & 26012 & 156608 &  929700   \\
 {\rm dim}($T_3$) 
            &    &    &  1 &  16 &  160 &  1304 &  9520 &  65056 &  426000   \\
 {\rm dim}($T_4$) 
            &    &    &    &   1 &   20 &   240 &  2268 &  18688 &  141156
 
 \end{tabular}
 \caption{Dimensions of the various blocks $T_Q$ of the transfer matrix, as functions of the strip width $L$.}
\label{tab:tmsize}
\end{table}

The values of ${\rm dim}(T_Q)$ can easily be obtained analytically
using generating function techniques. The result is that ${\rm
  dim}(T_Q)$ for a given (even) value of $L$ is the coefficient in the
term $q^Q$ in the polynomial expansion of
\begin{eqnarray}
 & \left( \frac{1+4q+q^2 + (1+q) \sqrt{1+6q+q^2}}{2q} \right)^{L/2} & + \nonumber \\
 & \left( \frac{1+4q+q^2 - (1+q) \sqrt{1+6q+q^2}}{2q} \right)^{L/2} & \,.
\end{eqnarray}
The dimension ${\rm dim}(T) = \sum_{Q=-L/2}^{L/2} {\rm dim}(T_Q)$ of
the total TM is then simply
\begin{equation}
  {\rm dim}(T) = (1+\sqrt2)^L + (1-\sqrt2)^L \,.
 \label{dimT}
\end{equation}
This is also the dimension of the (unique block of the) TM when monomers are allowed; see Sec.~\ref{sec:monomers} below.

\section{Columnar order at low temperature}
\label{sec:columnar}

\subsection{Possible crystalline orderings}

We present here MC results concerning the nature of the low-$T$ phase.
From the energy form, we expect at low-$T$ a proliferation of plaquettes
containing parallel dimers. A natural expectation is to have a single low-$T$
phase breaking the same symmetries as the ground-states in
Fig.~\ref{fig:GS}: we refer to such an order breaking both translation and
$\pi/2$ rotation symmetries as {\it columnar order}. On the other hand, from our knowledge of
QDM, we know that another type of order could also be stabilized: {\it
  plaquette order}. We describe more precisely this order below. As to
discriminate which kind(s) of phase(s) is (are) found at low-$T$ in the
dimer model Eq.~(\ref{eq:model}), we will introduce three different order
parameters. 

\subsubsection{Description of plaquette ordering} 
\label{sec:plaquette}

The QDM on the square lattice is believed to have some {\it plaquette}
long-ranged order  in some finite  region of parameter space at $T=0$.
In such a symmetry broken phase, one quarter  of the square plaquettes
are  spontaneously selected to  host a  pair of (quantum-mechanically)
resonating dimers.  The resulting  state breaks translation invariance
but is  invariant under $\pi/2$ rotation with  respect to the center of
any plaquette (see Ref.~\onlinecite{Leung} for an illustration).
In the quantum system, a plaquette phase has a (slightly) higher potential energy
than a columnar crystal, but the  stronger dimer {\it resonances}  lower the  plaquette  state energy
through the {\it kinetic} terms of the quantum Hamiltonian.
Of course, in our classical model, kinetic terms are
absent. Still, the  {\it thermal} fluctuations  of  the dimer locations around
each     ``flippable'' plaquette allow  to   gain   some {\it entropy}
(compared to that  of a columnar crystal) and  lower the free  energy.
The competition between entropy and  potential energy in the classical
system is analogous to that between kinetic and potential terms in the
QDM.  As the plaquette phase is likely to be realized  at $T=0$ in the
QDM,  it is    {\it a priori}  also  a  natural  candidate in   the  (finite
temperature) phase diagram of the classical model.

\begin{figure}
\begin{center}
\includegraphics[height=3cm]{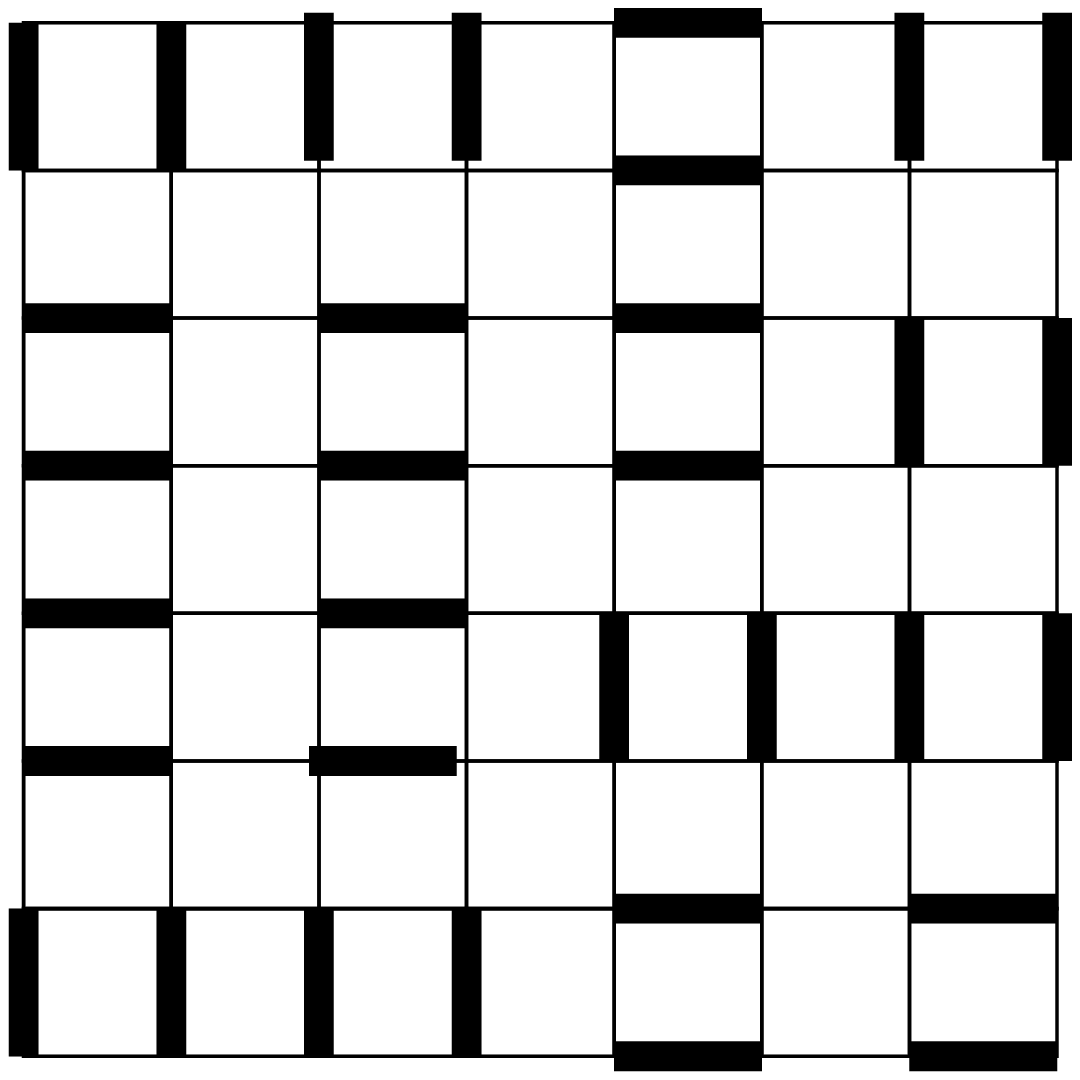}
\caption{A typical plaquette configuration.}
\label{fig:plaquette}
\end{center}
\end{figure}

In  a plaquette phase,  two  distant flippable  plaquettes are  almost
uncorrelated: if the first   one is dimerized, say,  horizontally, the
second can  be found in both states  with equal probability (hence the
$\pi/2$ rotation  symmetry).  This is    not  true for nearby  -   and
necessarily correlated    - plaquettes,  thus  defining   some  finite
correlation length.    A columnar state  can be  viewed as a plaquette
phase  in which this plaquette-plaquette  correlation length has grown
to infinity  so that all the  plaquettes of the lattice simultaneously
adopt the same orientation. A     typical      plaquette     configuration  is    displayed     in
Fig.~\ref{fig:plaquette}.     The plaquette phase breaks translational
symmetry but not $\pi/2$-rotational symmetry. A possible  scenario (eventually ruled  out  by the numerical results,
see below) could therefore be melting  of the columnar crystal through
an intermediate  plaquette  phase with  partial restoration   of  the
rotation symmetry.

\subsubsection{Order parameters}

{\it Complex columnar order parameter --- } We first use the definition (proposed in Ref.~\onlinecite{Sachdev}) of a complex columnar order parameter $\Psi_{\rm col}({\bf r})$ at site {\bf r}

\begin{eqnarray}
\Psi_{\rm col}({\bf r}) & = & (-)^{r_x} [ \hat{n}({\bf r}+{\bf x}/2)- \hat{n}({\bf r}-{\bf x}/2) ] \nonumber \\
&  + i & (-)^{r_y}  [ \hat{n}({\bf r}+{\bf y}/2)- \hat{n}({\bf r}-{\bf y}/2) ],
\label{eq:ColOP}
\end{eqnarray}
where ${\bf x,y}$ are unit vectors, and $\hat{n}$ is the dimer bond
occupation number ({\it i.e.} $\hat{n}({\bf r}+{\bf x}/2)$ is $1$ if there's
a dimer between site ${\bf r}$ and site ${\bf r}+{\bf x}$). We define the
associated columnar susceptibility as
\begin{equation}
\chi_{\rm col} = \frac{4}{L^2} ( \langle | \sum_{{\bf r} \in A} \Psi_{\rm col}({\bf r}) |^2 \rangle  -  \langle | \sum_{{\bf r} \in A} \Psi
_{\rm col}({\bf r}) | \rangle ^2),
\label{eq:ColChi}
\end{equation}
where the sums are taken only over the sublattice A. Another interesting quantity is the columnar Binder~\cite{Binder} cumulant 
\begin{equation}
B_{\rm col}=1-\frac{\langle|\Psi|^4\rangle }{2\langle|\Psi|^2\rangle ^2}.
\label{eq:ColBC}
\end{equation}
This Binder cumulant saturates to $1/2$ for a long-range ordered phase, and
scales to $0$ in the thermodynamic limit for a phase with no long-range
order, due to the Gaussian nature of the fluctuations of this order parameter. 
As was already noted in Ref.~\onlinecite{Leung}, this order parameter (and
associated quantities) is sensitive to translation symmetry breaking and
a non-zero expectation value detects both columnar and plaquette ordering.
Looking at the {\it phase} of $\Psi_{\rm col}$ can in principle discriminate between the two
phases: however the phase turns out to be a noisy observable in our simulations and
has no practical use. We will therefore use other indicators.

{\it  Dimer rotation  symmetry breaking   ---  } At sufficiently high  temperatures,   the system is  symmetric   under
$\pi/2$ rotations  so that the  average  number of  vertical dimers is
equal to the  average number of horizontal  ones. This also holds in a
plaquette phase, but is no longer true for a columnar state.
A convenient way of monitoring the $\pi/2$-rotation symmetry~\cite{Leung} is the 
Dimer Symmetry Breaking:

\begin{equation}
\label{eq:dsb}
D = 2/L^2 \left| N^c( \hdimera ) -  N^c(\vdimera ) \right|
\end{equation}
where $N^c(\hdimera )$ (resp. $N^c(\vdimera)$) is the number of horizontal
(resp. vertical) dimers in the configuration $c$. Normalization is
such that ${\rm D}=1$ in the pure columnar states of
Fig.~\ref{fig:GS}. It is also useful to define the corresponding susceptibility
$\chi_D=L^2(\langle D^2\rangle -\langle D\rangle^2)$ and Binder cumulant $B_D=1-\langle D^4\rangle /(3
\langle D^2\rangle ^2)$. 

{\it Plaquette order parameter --- } To discriminate {\it positively} the plaquette phase, we also use the following plaquette order parameter 
\begin{equation}
P=2/L^2 | \sum_{{\bf p}} (-)^{p_x+p_y} v_{\bf p} |
\label{eq:plaquette}
\end{equation}
where the sum is over all {\it plaquettes} of the lattice with coordinates
${\bf p}=(p_x,p_y)$, and $v_{\bf p}=1$ if the plaquette with coordinates
${\bf p}$ contains two parallel dimers ($v_{\bf p}=0$ otherwise). This
quantity can also be seen as a generalized energy at wave vector $(\pi ,\pi
)$. The staggered factor $(-)^{p_x+p_y}$ is essentially constant in a pure
plaquette state, and makes the sum vanish in a columnar state. The
expectation value $\langle P\rangle $ of the plaquette order parameter is
then $0$ in the columnar phase, and saturates to a finite value in the
thermodynamic limit in a plaquette phase. The associated plaquette
susceptibility is $\chi_P=L^2(\langle P^2\rangle -\langle P\rangle ^2)$ and
plaquette Binder cumulant $B_P=1-\langle P^4\rangle /(3
\langle P^2\rangle ^2)$. 

\subsection{Numerical results}

\subsubsection{Dimer rotation symmetry breaking}
\label{sec:DSBOP}

\begin{figure}
\begin{center}
\includegraphics[width=8cm]{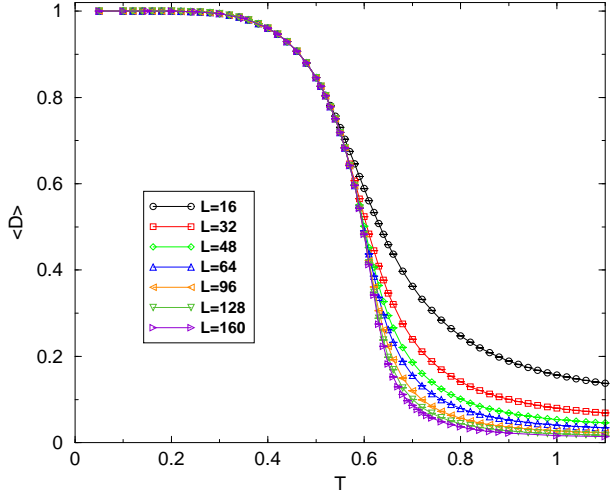}
\caption{(Color online) Dimer symmetry breaking order parameter $\langle D\rangle $ versus temperature $T$ for different system sizes.}
\label{fig:DSBOP}
\end{center}
\end{figure}

The expectation value $\langle D\rangle$ is displayed versus $T$ in
Fig.~\ref{fig:DSBOP} and clearly saturates to its maximum values at low $T$. 
The curves for different system sizes start to differ at a temperature
around $T\sim 0.6$ and in order to detect more finely the critical temperature
$T_c$, we use the corresponding susceptibility $\chi_D$ and Binder cumulant $B_D$.

\begin{figure}
\begin{center}
\includegraphics[width=8cm]{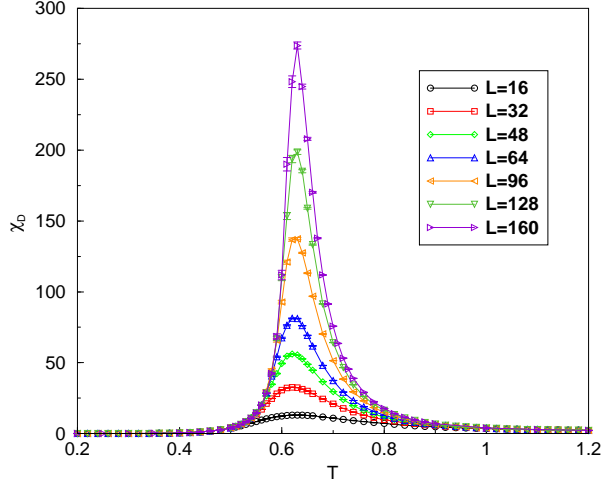}
\caption{(Color online) Dimer symmetry breaking susceptibility $\chi_D$ versus temperature $T$ for different system sizes.}
\label{fig:DSBChi}
\end{center}
\end{figure}

$\chi_D$ shows a pronounced peak around $T\sim 0.63$ (see
Fig.~\ref{fig:DSBChi}), signaling the onset of long-range order. Noticing that the temperature at which the susceptibility peaks
slightly drifts when increasing system size, we differ an estimation of
$T_c$ in favor of the Binder cumulant, which is known to allow accurate
determinations of $T_c$.

\begin{figure}
\begin{center}
\includegraphics[width=8cm]{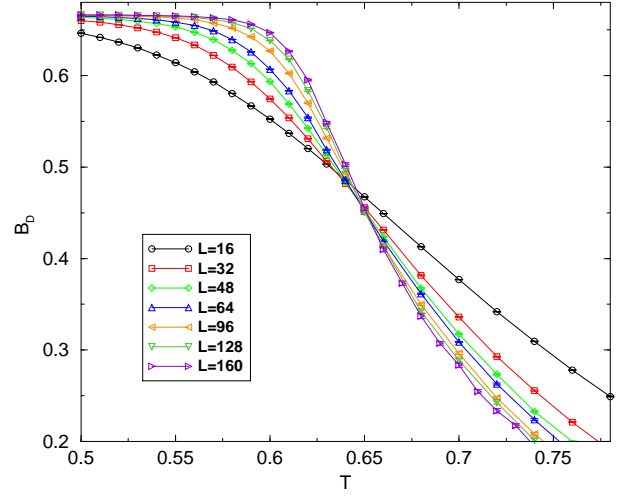}
\caption{(Color online) Dimer symmetry breaking Binder cumulant $B_D$ versus temperature $T$ for different system sizes.}
\label{fig:DSBBC}
\end{center}
\end{figure}

The Binder cumulant $B_D$ saturates in the thermodynamic
limit to $2/3$ at low $T$ (see Fig.~\ref{fig:DSBBC}) and we observe a
crossing of the curves for different system sizes for both cumulants at a
unique temperature $T_c$, signaling the entrance into the low $T$ columnar
phase. The critical temperature is estimated from this curve to be
$T_c=0.65(1)$.  The results of this section also indicate that plaquette
order is not present below $T_c$, but leave open the possibility of a plaquette phase at higher $T$.

\subsubsection{Plaquette correlations}
\label{sec:plaqcor}

\begin{figure}
\begin{center}
\includegraphics[width=8cm]{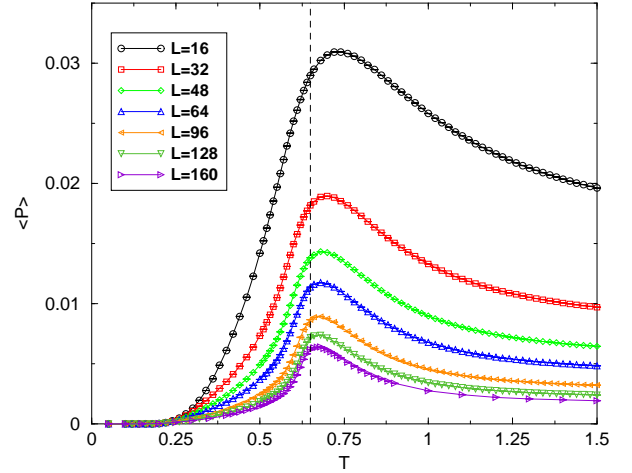}
\caption{(Color online) Plaquette order parameter $\langle P\rangle $ versus
  temperature $T$ for different system sizes. The dashed line denotes
  $T_c=0.65(1)$ as estimated by the Dimer symmetry breaking Binder cumulant.}
\label{fig:plaquetteOP}
\end{center}
\end{figure}

\begin{figure}
\begin{center}
\includegraphics[width=8cm]{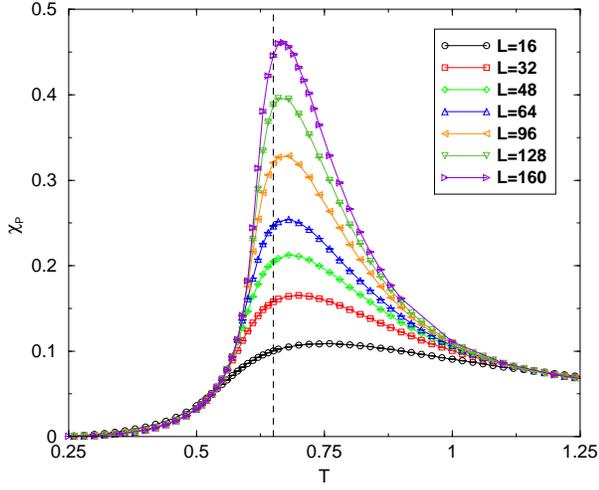}
\caption{(Color online) Plaquette susceptibility $\chi_P$ versus temperature
  $T$ for different system sizes.  The dashed line denotes $T_c=0.65(1)$.}
\label{fig:plaquetteChi}
\end{center}
\end{figure}

\begin{figure}
\begin{center}
\includegraphics[width=8cm]{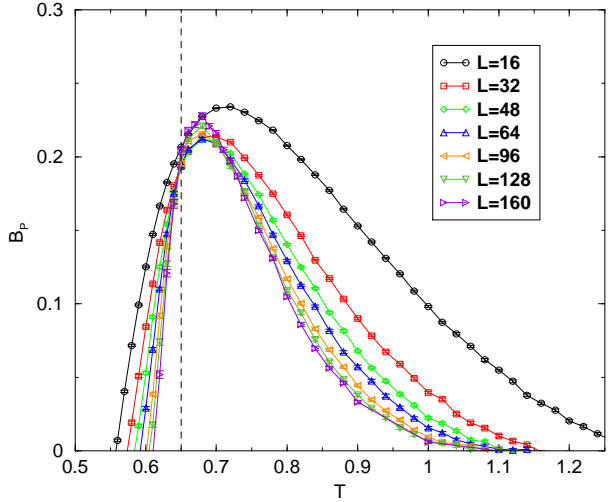}
\caption{(Color online) Plaquette Binder cumulant $B_P$ versus temperature
  $T$ for different system sizes. The dashed line denotes $T_c=0.65(1)$.}
\label{fig:plaquetteBC}
\end{center}
\end{figure}

The expectation value of the plaquette order parameter $\langle P\rangle $
shows a non-monotonous behavior as a function of $T$ (see
Fig.~\ref{fig:plaquetteOP}), with an order parameter peaking close to
$T_c \sim 0.65$ {\it from above} for all system sizes. One also immediately
notes that $\langle P\rangle $ has overall small values and {\it decreases}
  with system size.  The plaquette susceptibility $\chi_P$ peaks slightly
above $T_c$ (see Fig.~\ref{fig:plaquetteChi}): we interpret this as
plaquette correlations being the strongest just before the entrance into the
columnar phase. Even though the $\chi_P$ values are very small values as
compared to other typical susceptibilities (see for example
Fig.~\ref{fig:DSBChi}), long-range plaquette order could survive in the
thermodynamic limit. This is clearly ruled out by the behaviour of the
plaquette Binder cumulant (see Fig.~\ref{fig:plaquetteBC}) which is
non-monotonous as well: $B_P$ starts to rise from its high-$T$ zero value when
decreasing temperature and suddenly drops down to zero at a temperature slightly above $T_c$. This excludes long-range
plaquette order.

We conclude that (strong) plaquette correlations are present, start to develop as one
decreases $T$, peak just above $T_c$, but do not form a true thermodynamic
phase as they are suddenly overtaken by columnar order. However, as we show
in the next section, these correlations possibly ``pollute'' the finite-size
behavior of the complex columnar order parameter.

\subsubsection{Complex columnar order parameter}
\label{sec:ColOP}

\begin{figure}
\begin{center}
\includegraphics[width=8cm]{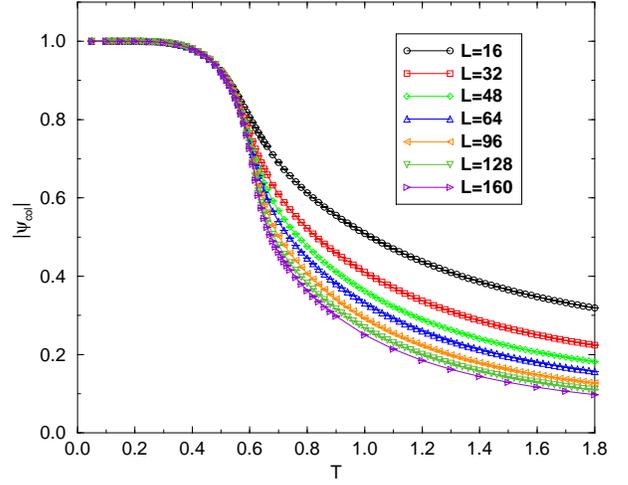}
\caption{(Color online) Columnar order parameter $\langle | \Psi_{\rm col} |
  \rangle$ as a function of temperature $T$ for different system sizes.}
\label{fig:ColOP}
\end{center}
\end{figure}

\begin{figure}
\begin{center}
\includegraphics[width=8cm]{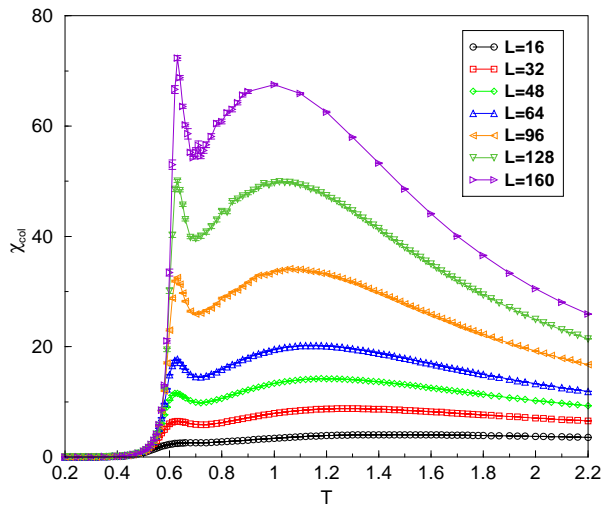}
\caption{(Color online) Columnar susceptibility $\chi_{\rm col}$ as a
  function of temperature $T$ for different system sizes.}
\label{fig:ColChi}
\end{center}
\end{figure}

\begin{figure}
\begin{center}
\includegraphics[width=8cm]{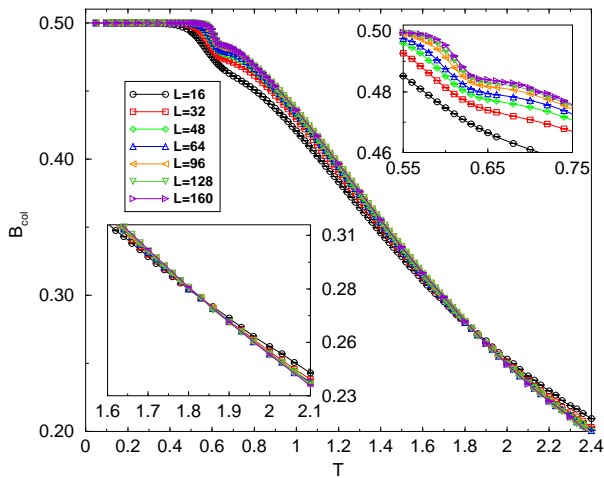}
\caption{(Color online) Columnar Binder cumulant $B_{\rm col}$ as a function
  of temperature $T$ for different system sizes. Right inset: Anomaly near $T\sim 0.63$. Left inset: Pseudo-crossing around $T\sim 1.8$.}
\label{fig:ColBC}
\end{center}
\end{figure}

The columnar order parameter $\langle | \Psi_{\rm col} | \rangle =
\frac{2}{L^2} \langle |
\sum_{{\bf r} \in A} \Psi_{\rm col}({\bf r}) | \rangle$ is displayed in
Fig.~\ref{fig:ColOP} for different system sizes (from $L=16$ to
$L=160$). One clearly sees order setting in at low $T$, and the
curves for different $L$ start to separate roughly around $T\sim0.6$. To
determine more precisely the critical point, we inspect the behavior of the
columnar susceptibility (Eq.~(\ref{eq:ColChi}) and Fig.~\ref{fig:ColChi})
and Binder cumulant (Eq.~(\ref{eq:ColBC}) and Fig.~\ref{fig:ColBC}).

The columnar susceptibility has an unexpected behavior. It exhibits two
peaks: the first is quite sharp and localized around $T\sim 0.63$, and the
second is much broader around $T\sim 1$. It is also to be noted that whereas
for small system sizes ($L\leq 96$), the first peak is smaller than the
second one, the tendency is inverted for the two largest system sizes. This
could mean that the second peak actually saturates to a finite value in the thermodynamic
limit. Another possible scenario is that the two
peaks merge, which is not unlikely noticing that the positions of the
maximum of the second peaks shift toward lower $T$ when increasing
$L$. Unfortunately, the currently available system sizes do not allow us to
draw definitive conclusions on the scaling behaviour of $\chi_{\rm col}$. 

The columnar Binder cumulant (Fig.~\ref{fig:ColBC}) also displays unusual
features: a very flat pseudo-crossing of the curves for
different $L$ is observed around $T\sim1.8$ (see zoom on left inset of
Fig.~\ref{fig:ColBC}) and a marked anomaly around $T\sim0.63$ (see zoom on
right inset). A crossing of curves corresponding to different system sizes
for a Binder cumulant usually denotes a transition to a long-range ordered
phase. However, the crossing observed here is very flat and it is actually
almost impossible within the statistical accuracy to locate a single
crossing point for the three largest samples. The anomaly around $T\sim
0.63$ is particularly singular and we are not aware of such a behavior
being reported for a Binder cumulant in the literature. The fact that two
singularities are observed both in $\chi_{\rm col}$ and $B_{\rm col}$ could be interpreted at first glance as signs of an
intermediate phase. However, given our previous findings of strong but no
long-range ordered plaquette correlations, it appears likely that the
second feature at high $T$ actually disappears in the thermodynamic
limit, whereas the first one around $T\sim0.63$ subsists. Indeed, whereas both the columnar
susceptibility and Binder cumulant are marked at the lowest temperature, the
temperature at which $\chi_{\rm col}$ peaks is different from the one at which $B_{\rm col}$ shows a crossing.

We believe that the plaquette correlations found in Sec.~\ref{sec:plaqcor} affect the
finite size behavior of other observables and are in particular responsible
for the behaviors observed in the columnar susceptibility and Binder cumulant. The fact
that strictly speaking, $\chi_P$ does not peak where
the second peak in $\chi_{\rm col}$ is present (even though the latter
drifts with system size) might indicate that other types of correlations are also present above $T_c$.

In conclusion of this section, we find that the model Eq.~(\ref{eq:model}) shows a unique
transition to a columnar order at $T_c=0.65(1)$ with no intermediate phase,
but with strong plaquettes correlations above $T_c$.

\section{Nature of the phase transition}
\label{sec:KT}

The model Eq.~\ref{eq:model} displaying a phase transition at $T_c=0.65(1)$ separating a low $T$ columnar phase from a high $T$ phase (that we will describe more carefully in Sec.~\ref{sec:highT}), we now investigate the nature of this phase transition.

\subsection{Energy cumulant}

\begin{figure}
\begin{center}
\includegraphics[width=8cm]{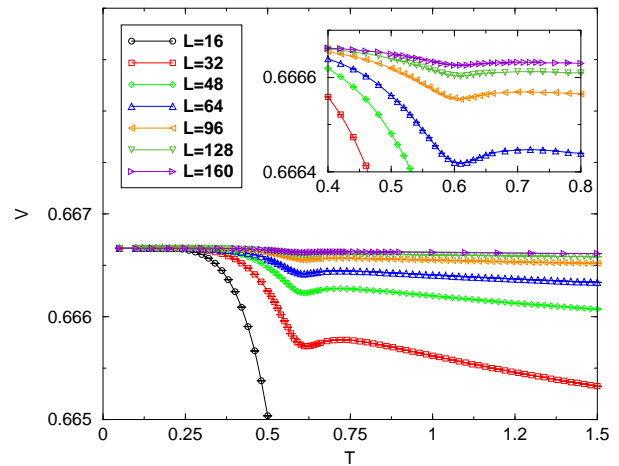}
\caption{(Color online) Energy cumulant $V$ versus temperature $T$ for different system sizes. Inset: zoom on the large size samples.}
\label{fig:Ecum}
\end{center}
\end{figure}

We first plot in Fig.~\ref{fig:Ecum} the behavior versus temperature $T$ of the energy cumulant~\cite{Challa} :
\begin{equation}
\label{eq:Ecum}
V=1-\frac{\langle E^4\rangle }{3\langle E^2\rangle ^2}.
\end{equation}
In both a disordered and in an ordered phase, this cumulant saturates to
$2/3$ in the thermodynamic limit~\cite{Challa}. This is also the case at the
critical point for a second order phase transition (even though the energy
distribution is not Gaussian), whereas for a first order transition, it
admits a non-trivial minimum (different from $2/3$) in the thermodynamic
limit~\cite{Challa}. We are not aware of any predictions for a KT
transition. 

Our results for $V(T)$ show a clear dip close to the critical temperature
$T_c=0.65(1)$ for all system sizes, but this minimum scales to $2/3$ in the
thermodynamic limit as can be clearly seen in the inset of
Fig.~\ref{fig:Ecum}. These results {\it exclude} a first order
transition. 

\subsection{Specific Heat}

We now consider the second cumulant of the energy, {\it i.e.} the specific heat per site, defined as :
\begin{equation}
\label{eq:Cv}
\frac{C_v}{N}=\frac{1}{N}\frac{\langle E^2\rangle -\langle E\rangle ^2}{T^2}.
\end{equation}

\begin{figure}
\begin{center}
\includegraphics[width=8cm]{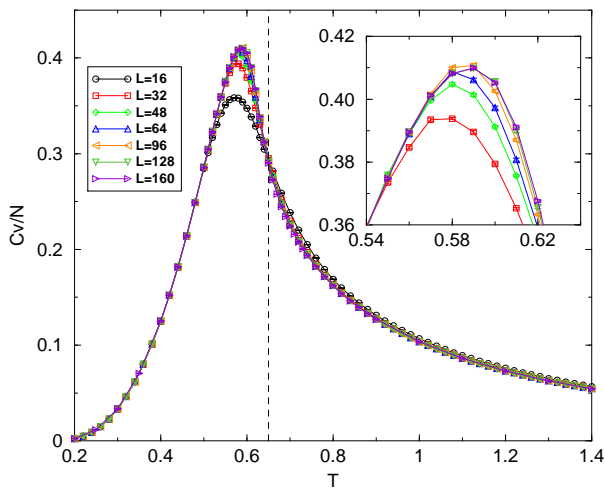}
\caption{(Color online) Specific heat per site $C_v/N$ versus temperature $T$ for different system sizes. The dashed line indicates $T_c=0.65(1)$. Inset: zoom on the specific heat peak.}
\label{fig:Cv}
\end{center}
\end{figure}

The specific heat per site shows a pronounced peak which does {\it not}
diverge in the thermodynamic limit (see Fig.~\ref{fig:Cv} and its inset). We
can see that this peak is located at a temperature $T\sim 0.59$
different from the critical temperature $T_c=0.65(1)$, denoted by a dashed
line in Fig.~\ref{fig:Cv}. For a second-order phase transition, we would
have expected either a divergence of the specific heat at $T_c$ (if the
critical exponent $\alpha>0$) or a cusp (for $\alpha<0$). Our results show
that $C_v$ is completely featureless at $T_c$: this is typical of a {\it
  Kosterlitz-Thouless transition}. We will see in the following sections
that this is naturally expected noticing the nature of the high-temperature
phase that we now address.

\section{High-temperature phase}
\label{sec:highT}

It was demonstrated more than 40 years ago that the $T=\infty$ point of our
model (which is the classical dimer covering of the square lattice) is
critical: the system possesses correlations functions that decay
algebraically with distance~\cite{FisherStephenson}. In this section we
consider the fate of various correlation functions in the whole
high-$T$ region $]T_c,\infty[$. 

\subsection{Dimer-Dimer correlation functions}

We have calculated two types of dimer-dimer correlation functions in the MC
simulations. Both types concern dimers which are chosen for simplicity in
the same orientation. We take horizontal dimers without loss of generality. The first correlator which we dub ``longitudinal'' is the connected correlation function of two horizontal dimers on the same row separated by a distance $x$ : 
\begin{equation}
\label{eq:DDhh}
 G^l(x)= \langle \hat{n}_{\hdimerb}({\bf r})\hat{n}_{\hdimerb}({\bf r}+(x,0)) \rangle -1/16,
\end{equation}
where $\hat{n}_{\hdimerb}({\bf r})=1$ for a horizontal dimer at site {\bf r}, and  $0$ otherwise. The constant $1/16$ stands for the dimer density squared. The second one is the ``transverse'' correlation function of two dimers separated by a distance $x$ on the same column :
\begin{equation}
\label{eq:DDhv}
 G^t(x)= \langle \hat{n}_{\hdimerb}({\bf r})\hat{n}_{\hdimerb}({\bf r}+(0,x)) \rangle -1/16.
\end{equation}
At $T=\infty$, the exact calculations of Ref.~\onlinecite{FisherStephenson} give the asymptotic results
\begin{equation}
G^l(x)\sim (-)^x \frac{1}{\pi^2 x^2} + O(x^{-3})
\end{equation}
 and  
\begin{eqnarray}
\label{eq:ddcorr.texact}
G^{t}(x) & \sim & \frac{1}{\pi^2x^2} + O(x^{-3}) \quad \quad \quad \textrm{x odd} \\
& \sim & -\frac{1}{\pi^2x^4} + O(x^{-6})  \qquad \textrm{x even}
\label{eq:ddcorr.texact2}
\end{eqnarray}

\begin{figure}
\begin{center}
\includegraphics[width=8cm]{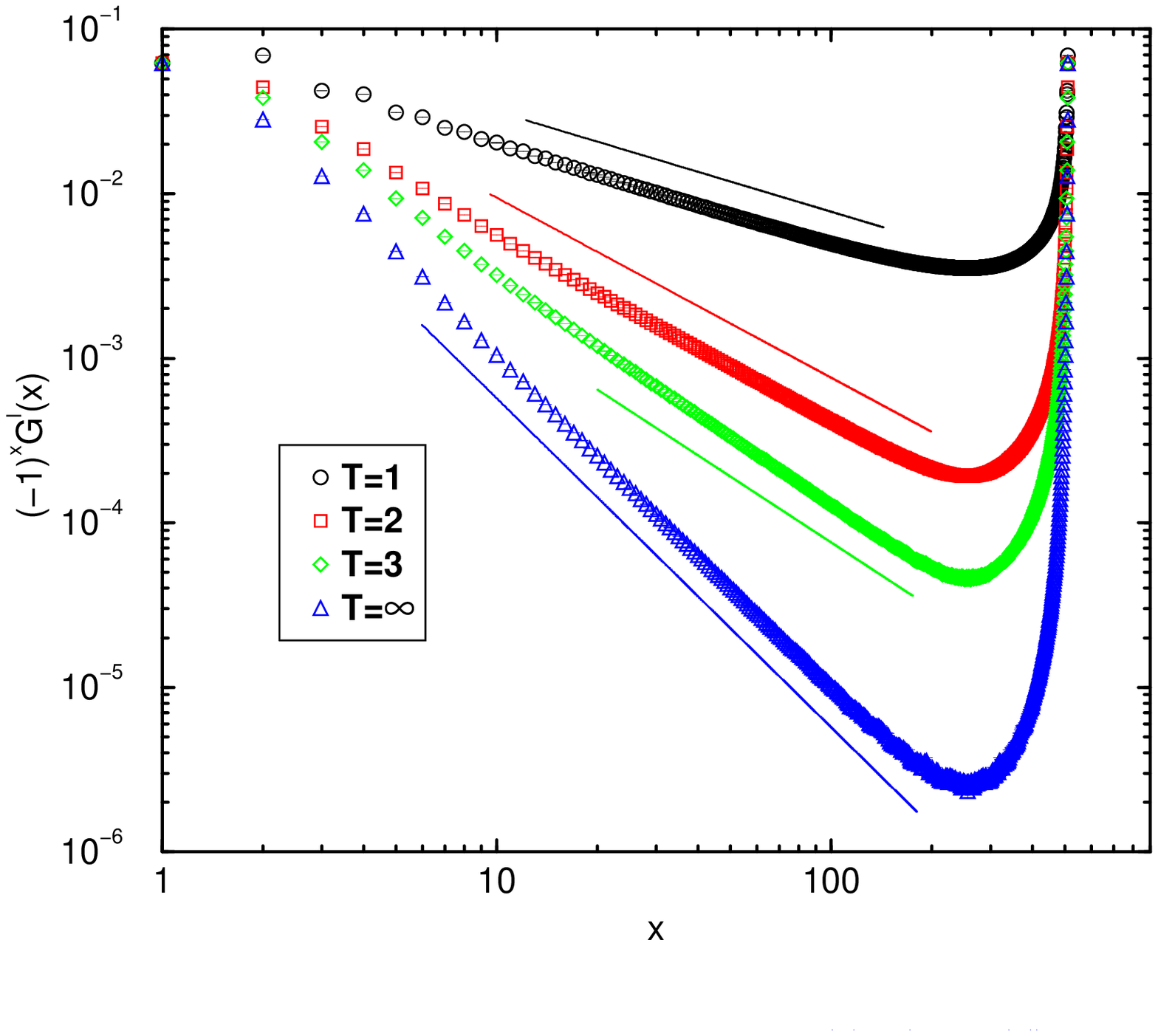}
\caption{(Color online) Staggered longitudinal dimer-dimer correlation
  function $(-)^xG^l(x)$ versus distance $x$ between dimers for four
  different temperatures $T$ (log-log scale). The straight lines correspond to decay exponents $\alpha_d(T)$ calculated in  Sec.~\ref{sec:theo}.}
\label{fig:DDhh}
\end{center}
\end{figure}

For all finite $T\geq T_c$, we find that the longitudinal correlation
function $G^l(x)$ remains staggered, and that it {\it decays algebraically,
  with a decay exponent $\alpha_d$ that varies continuously with the
  temperature $T$:}
\begin{equation}
G^l(x)\sim (-)^x A(T) x^{-\alpha_d(T)}
\label{eq:DDhh2}
\end{equation}
for large $x$ (with $A(T)$ an amplitude). In Fig.~\ref{fig:DDhh}, we
represent  $(-)^xG^l(x)$ for four different $T=1,2,3$ and $T=\infty$ on a
log-log scale to emphasize the power-law decay. The algebraic decays are
eventually cut around $L/2$ due to the PBC (system size is here
$L=512$). The value of the decay exponent $\alpha_d(T)$ can be estimated 
from these plots, however the symmetry around $L/2$ due to
the PBC makes a high-precision determination of the exponent difficult, since
it would depend on the range of distances used in the fit. We will use
alternative methods (TM calculations and winding fluctuations)
in Sec.~\ref{sec:theo} to estimates decay exponents. To show that the
different ways of estimating the decay exponents are consistent, we have
plotted in Fig.~\ref{fig:DDhh} lines corresponding to the decay exponents
found in Sec.~\ref{sec:theo} (at $T=\infty$, we take the exact result $\alpha_d(T=\infty)=2$). These lines are
in perfect agreement with the first part of the correlation function
$(-)^xG^l(x)$ which is not affected by the periodicity.

\begin{figure}
\begin{center}
\includegraphics[width=8cm]{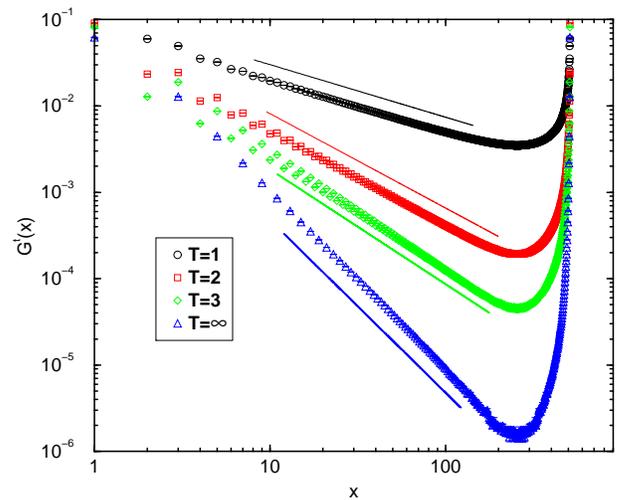}
\caption{(Color online) Transverse dimer-dimer correlation function $G^t(x)$
  versus distance $x$ between dimers for four different temperatures $T$ (log-log scale). The straight lines correspond to decay exponents $\alpha_d(T)$ calculated in Sec.~\ref{sec:theo}. For the $T=\infty$ curve, the results for even $x$ have been omitted (as they are negative - see Eq.~\ref{eq:ddcorr.texact2}).}
\label{fig:DDhv}
\end{center}
\end{figure}

In Fig.~\ref{fig:DDhv}, we plot the transverse correlation function $G^t(x)$
for the same $T$. We also find here a power-law decay with the same
exponent $\alpha_d(T)$ as for the longitudinal correlation function
($(-)^xG^l(x)$ and $G^t(x)$ essentially coincide for large $x$). Small
deviations can however been found at small distances (see $x<10$ in
Fig.~\ref{fig:DDhv}), where the data for odd or even $x$ do not exactly
coincide. This odd/even distinction is already present in the $T=\infty$
case - see Eq.~(\ref{eq:ddcorr.texact}) and~(\ref{eq:ddcorr.texact2}). We find that $G^t(x)$ is well fitted
by the expression:

\begin{eqnarray}
G^{t}(x) & \sim & A(T)  x^{-\alpha_d(T)} \qquad \quad \quad \quad \quad \quad \quad \textrm{\rm x odd} \\
& \sim & A(T)  x^{-\alpha_d(T)} + B(T) x^{-\omega(T)} \quad \textrm{\rm x even}
\label{eq:DDvv2}
\end{eqnarray}

where $A(T)$ is the same amplitude as the one found for the longitudinal correlation function, $B(T)$ a negative constant and $\omega(T)$ a subleading correction exponent for even $x$. Our data are compatible with $\omega(T)\sim 2$. 

\subsection{Monomer-Monomer correlation function}

Besides dimer-dimer correlation functions, it is useful to consider the
correlator between {\it monomers}, which are defined as sites not paired to
any neighboring site by a dimer. Monomers are by definition absent in the
model, however we can calculate the properties of two test monomers in a
background of dimers by considering the monomer-monomer
correlation function~\cite{FisherStephenson,Krauth,Sandvik}
\begin{equation}
\label{eq:Monomer}
M({\bf x})\sim Z({\bf x}),
\end{equation}
where $Z({\bf x})$ denotes the number of possible configurations where the
two test monomers are separated by a vector ${\bf x}$. Note that this is
exactly the correlation function $C_{Q_0}(M)$ defined in the TM
section~\ref{sec:TM} (we changed notations from $M$ to ${\bf x}$ and
$C_{Q_0}$ to $M$ for consistency with previous works).  For a bipartite
lattice such as the square lattice, $M({\bf x})=0$ if monomers are located
on the same sublattice. From now on, we will consider only monomers on opposite sublattices.
To simplify calculations, we focus on the case ${\bf x}=(x,0)$ (two monomers on the same row). The
proportionality constant is taken such that
$M(1)=1$~\cite{FisherStephenson,Sandvik}. 

\begin{figure}
\begin{center}
\includegraphics[width=8cm]{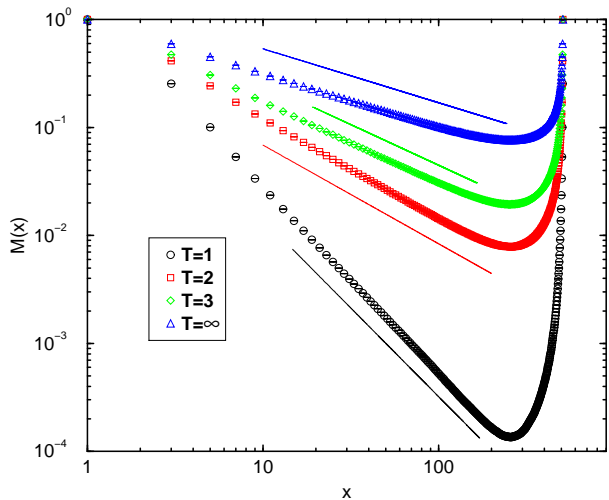}
\caption{(Color online) Monomer-monomer correlation function $M(x)$ versus
  distance $x$ between monomers for four different temperatures ($T$ log-log scale). The straight lines correspond to decay exponents $\alpha_m(T)$ calculated in Sec.~\ref{sec:theo}. For the $T=\infty$ curve, we used the exact result $\alpha_m(T=\infty)=1/2$.}
\label{fig:Monomer}
\end{center}
\end{figure}

At $T=\infty$, the exact result $M(x)\sim x^{-1/2}$ holds in the thermodynamic limit~\cite{FisherStephenson}. At finite $T$, we can estimate $M(x)$ with high-precision thanks to the worm algorithm (see Sec.~\ref{sec:MC}). Results for $M(x)$ are displayed versus $x$ on a log-log scale in Fig.~\ref{fig:Monomer} for the four temperatures used for the dimer-dimer correlation functions: we also observe here power-laws, with an exponent $\alpha_m(T)$ that appears to vary continuously with $T$:
\begin{equation}
\label{eq:MonomerDecay}
M(x) \sim x^{-\alpha_m(T)}.
\end{equation}
For the same technical reasons as for the dimer-dimer correlation functions,
we do not estimate directly from this plot the values of $\alpha_m(T)$, as
the error bars would be too large. We will obtain precise estimates of
$\alpha_m(T)$ in the next section~\ref{sec:theo}, which we already display
as straight lines in Fig.~\ref{fig:Monomer} to show the good agreement with
the real-space measures of $M(x)$. As will be understood, we note
that whereas $\alpha_d(T)$ decreases when lowering $T$, $\alpha_m(T)$
increases (this can be clearly seen with the identical color chart of
figures~\ref{fig:DDhh},~\ref{fig:DDhv} and~\ref{fig:Monomer}).

\section{Theoretical interpretation}
\label{sec:theo}

We recapitulate the MC results on the finite-$T$ behavior of the interacting dimer model Eq.~(\ref{eq:model}) with $v<0$: we observe a transition from a high-$T$ critical phase with continuously varying critical exponents to a low-$T$ crystalline phase of dimers. Even though other types of correlations are present (such as plaquette correlations), there is no intermediate phase in-between. The transition is found to be of KT type. The existence of the high-$T$ critical phase, the floating exponents, the KT
transition and the dimensionality $d=2$ of the problem naturally suggest
{\it Coulomb gas physics}~\cite{CG}. In the following, we will indeed
describe the (non-exact) mapping of the interacting dimer model to a CG that
will account for all the above findings. Moreover, thanks to results of CFT
and use of TM calculations, we will be able to determine with high-precision
the critical exponents $\alpha_d(T)$ and $\alpha_m(T)$, and the associated
CG constant. 

The following section will also have the advantage of reconciliating different
points of view on the model Eq.~(\ref{eq:model}). Readers familiar with height models will find an alternative
derivation of the effective height action for dimer configurations.
For readers familiar with CG physics, the TM calculations presented here
will allow a high precision test of the CG predictions, with a precision
probably not reachable in other models. For readers interested in the dimer
enumeration problem at $T=\infty$, an interesting relationship between dimer
winding numbers fluctuations and CG constant will be derived in
passing. Finally, for readers coming from the quantum condensed matter
community, these results have profound implications for the high-$T$ regime
of QDM, as will be discussed in Sec.~\ref{sec:discuss}.

\subsection{Mapping to a Coulomb gas}

\subsubsection{Height model}
\label{sec:height}

To obtain a CG picture of the interacting  dimer model, we first use a
{\it  height} description  of dimer configurations~\cite{Blote}. Each
plaquette  of the square lattice is  assigned a real-valued height $z$
in  the  following  way:   going counterclockwise  around  an   even
(resp. odd) site on  the square lattice,  $z$ changes by $+3/4$ if the
bond crossed is  occupied by a   dimer and by   $-1/4$ if it is  empty
(resp. $-3/4$ and $+1/4$). These units are  chosen such that a monomer
on the even (resp. odd) sublattice corresponds to a dislocation of $1$
(resp.  -1) in the  height. To fix the absolute  height, one fixes the
plaquette at the origin to have, say, $z(0)=0$. By  integrating out  the short distance   fluctuations  of $z(r)$, one
obtains  an effective action $S_{\rm  eff}[h]$  for the coarse-grained
height   $h(r)$, defined in the  continuum,   which corresponds to the
long-wavelength modes of $z(r)$.

{\it  Locality} --- The form  of this effective action is constrained
by the fact that the microscopic model is local in  terms of the dimer
degrees of freedom. Consider a finite area $A$  of the system and some
fixed coarse-grained height  $h(r)$ for $r\in  A$. The associated free
energy (obtained by  summing over all microscopic dimer configurations
compatible with  the given $h(r)$)   should  not depend  on the  dimer
positions {\em outside} $A$. However,  by shifting the dimers along a
closed  loop,   the dimer configuration  inside   is unchanged but the
microscopic  height $z(r)$ is uniformly shifted  by $+1$ (or $-1$) for
all the   plaquettes located inside  the  loop (and  so for the coarse
grained  height $h$).  Doing so for  a large loop surrounding $A$, one
therefore shows that $S_{\rm eff}$ must satisfy $S_{\rm eff}[h]=S_{\rm
eff}[h+1]$ for any physical $h$.

{\it $\pi/2$ rotation symmetry} --- Consider a large but finite square
area $A$ of  the lattice with linear even size $L$.  Outside $A$, the
dimers    are  assumed  not   to  cross  the   boundary  of  $A$.  Let
$z(r)=z_0+d(r)$ be the height  inside $A$ and  $z_0$ the height of the, say, 
lower left corner of $A$~\cite{note.corners}. Whatever the   dimer  locations inside  $A$
(compatible with the  constraint above), one  can move the dimers {\it
inside} $A$ by a  $\pi/2$ rotation $\mathcal   R$ with respect  to its
central plaquette.  $z$ is unchanged outside $A$ but  for $r\in A$ the
height is  changed  to $z'(r)=z_0-d(\mathcal{R}(r))$.  Now  we  take a
smooth height profile $h(r\in  A)$  and evaluate the  associated  free
energy $S_{\rm eff}[h]$ by summing  over all microscopic dimerizations
giving  the   same coarse grained    height $h$.  By   the rotation
described    above,        we       know    that  another       height
$h'(r)=-h(\mathcal{R}(r))$  corresponds     to the    same    set   of
dimerizations of $A$,  up to a rotation.   Because $\mathcal{R}$  is a
symmetry of the lattice and of the  dimer-dimer interactions, both $h$
and  $h'$  must  have  the  same free    energy   and we  get  $S_{\rm
eff}[h]=S_{\rm eff}[-h]$. Here we ignored boundary effects at the edge
of $A$, which should be negligible for a large enough area.

\begin{figure}
\includegraphics[width=8cm]{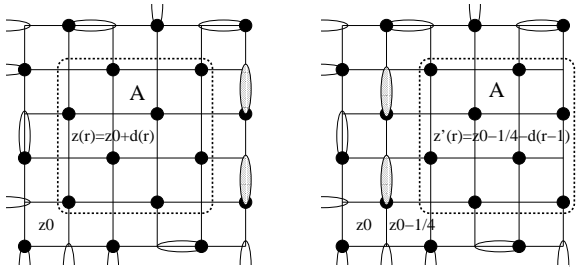}
\caption{
Assuming that the  column of  sites located  immediately  to the right
side of the area $A$ is occupied  by vertical dimers (shaded) and that
no dimer crosses the boundary of $A$  (dashed line), one can shift the
whole area by one  lattice spacing to the  right.  The vertical dimers
are then moved to the left  side. Such a ``local'' translation changes
the height   profile   inside    $A$   from  $z(r)=z_0+d(r)$     to
$z'(r)=z_0-\frac{1}{4}-d(r)$.  The change $d(r)\to-d(r)$ reflects  the
fact that the two  sublattices are exchanged  inside A.  $z_0$  is the height at the  lower left corner of $A$ before
the translation.}
\label{fig:height}
\end{figure}

{\it Translation  symmetry} --- In addition  to the assumption that no
dimer crosses the boundary of  $A$, we assume  that some dimers occupy
all the vertical bonds on the right side  of $A$ (shaded dimers in the
left panel of Fig.~\ref{fig:height}).  One can shift the dimers of $A$
by one lattice spacing to the right provided that the column of dimers
on the right  side of  $A$  is put back on  the  left side after  the
translation  (shaded    dimers    in      the    right  panel       of
Fig.~\ref{fig:height}).  If the  height inside $A$ is $z(r)=z_0+d(r)$,
the    new        height       after     the      translation     is
$z'(r)=z_0-\frac{1}{4}-d(r-1)$. As before, $z_0$ is  the height of the
lower left corner of $A$  before the move (and $z_0-\frac{1}{4}$ after
the translation).  Again, we evaluate the free energy $S_{\rm eff}[h]$
associated to a smooth height profile  by summing over all microscopic
dimerizations giving the coarse-grained height  $h$.  From the ``local
translation''      above,  one    shows    that    the   smooth height
$h'(r)=-\frac{1}{4}-h(r-1)\simeq-\frac{1}{4}-h(r)$ corresponds to  the
same set   of  dimerizations  of $A$,  but   shifted   by one  lattice
spacing. Because  of  the {\em translation  invariance}  of the model,
both  coarse-grained  height   profiles therefore  have  the same  free
energy: $S_{\rm eff}[h]=S_{\rm  eff}[-h-\frac{1}{4}]$.   As before  we
ignored boundary effects  at the edge   of $A$. We also neglected  the
variations  of $h$ at  the lattice spacing scale: $h(r)\simeq h(r+1)$.
This is natural for a smooth height, obtained  from a microscopic (and
discrete) $z(r)$ by filtering out short wavelength modes.

From  the discussion above, it appears  that lattice symmetries of the
dimer model  implies   not only  spatial symmetries for   the effective
height model but also some {\it periodicity} of the free energy in the
height space. A similar result was previously obtained by Henley and
coworkers using the concept of ``ideal states''~\cite{Henley}.

Combining the constraints of translation  and rotation symmetries,  we
get that $h\to -h$ and $h\to  h+\frac{1}{4}$ should keep the effective
action invariant. This shows that the only allowed ``potential'' terms
are $\cos(2\pi  p  h)$ with $p$  an integer  multiple of 4.    As only
long-wavelength modes  $(k\ll 1)$  are kept in a  coarse-grained height,
only terms    with  a   minimal   number  of  space  derivatives   are
important.   This naturally leads  to    an ``elastic'' term  $(\nabla
h)^2$.  We eventually   get a  sine-Gordon model   for  the coarse-grained
height:

\begin{equation}
\label{eq:eff}
S_{\rm eff}= \int  d{\bf r} \,  \left[ \pi  g |\nabla  h({\bf r})|^2 +
\sum_{p=4,8,12,\cdots} V_p\cos \big(2\pi p h({\bf r})\big) \right]
\end{equation}

So  far we only invoked symmetry  arguments to constrain the form of
the effective action.   It turns  out  that the  elastic and potential
terms with $p=4$ have a simple physical interpretation in terms of the
dimer model.  The  gradient term favors  ``flat'' heights.  There are
indeed more  dimer  configurations  corresponding  to  a  flat average
height than a tilted one. This comes from  the fact that a dimer shift
along a closed loop is possible only if  the height is constant (up to
small discretization   effect) along the  loop.   Thus,  there is more
room for dimer moves if the average slope is  small (in which case
there are many small ``iso-$h$'' loops available) than if the slope is
large (fewer ``iso-$h$'' curves). As for  the $\cos(8\pi h)$ terms, it
has   4  minima  ($h=\frac{1}{8},\frac{3}{8},\frac{5}{8},\frac{7}{8}$)
which  precisely  coincide with the average   height of the 4 columnar
configurations     (ground-states)  which minimize   the  dimer-dimer
interaction energy (see  Fig~\ref{fig:GS}).  The  model therefore describes a
combination of entropy and energy (locking) effects.

It is a standard result~\cite{Coleman} that in 2D the relevance in
the Renormalization Group (RG) sense of the cosine term depends on the
period and on the stiffness constant $g$~\cite{note.g}. The cosine
term is relevant (and locks the height) if $g>p^2/4$ whereas it
renormalizes to zero when $g<p^2/4$.  In the latter case the
long-distance theory is a free field (elastic term only) and the
system is critical (``rough'' in the height model terminology).  The
transition between the two phases is of the KT type. For completeness,
this classic RG calculation is reproduced in
Appendix~\ref{sec:rgsg}.

\subsubsection{Coulomb gas}
\label{sec:CG}

It  is well-known  that  the sine-Gordon  model  is  equivalent  to  a
(low-density)   one-component   CG~\cite{CG}. This  standard
mapping is reproduced in Appendix~\ref{sec:coulomb2sg}.  This point of
view has  the  advantage (over the  sine-Gordon  model described above)
that it provides a framework to  understand the role of the(here
suppressed) monomers  in the model,   as well  as other  operators  or
correlations.

In the mapping from the sine-Gordon model to a CG model, the height
field is conjugate to the {\em electric} charge density. Similarly, it
can be shown that magnetic charges correspond to topological defects
(dislocations) in the height. A dual magnetic charge $m=1$ (resp.
$m=-1$) corresponds to a dislocation in the height field, which can be
inserted ``by hand'' through the inclusion of monomers on the even
(resp. odd) sublattice or through appropriate boundary conditions (see
the discussion in Sec.~\ref{sec:TM}). The
exponent~\cite{note.exponent} associated to the insertion of a
particle with electromagnetic charge $(e,m)$ is given by~\cite{CG}
\begin{equation}
\label{eq:CG}
\alpha (e,m) = e^2 / g + g m^2,
\end{equation}
where $g$    is   the  {\it  Coulomb  gas     coupling constant}.  The
normalization of $g$ and of the  height field in Eq.~(\ref{eq:eff}) have been
chosen such that $e$ and $m$  are integers and that standard expressions~\cite{CG}
for $g$ are recovered.

In the CG picture, inserting an electric charge $e$ is implemented by
a vertex operator $V_e({\bf r})=:\exp\big(2i\pi e h({\bf r})\big):$
appearing in the Fourier expansion of any operator periodic in the
coarse-grained field.  Such a term with $e=1$ actually appears as a
continuum limit contribution in the definition of the dimer
operator~\cite{Fradkin}: this allows to identify the continuum limit
of the dimer number as an operator with electric charge $e=1$. To
magnetic charges (monomers) correspond dual operators $\tilde{h}({\bf
  r})$; however the fugacity for these magnetic charges is fixed to
zero as we consider close-packed dimers.  The dimer-dimer correlation
function decays with an exponent related to the dimension of the $e=1$
operator, $\alpha_d=\alpha (1,0)=1/g$, and the monomer-monomer
correlation function decays with the exponent $\alpha_m=\alpha
(0,1)=g$. This leads in particular to the prediction
$\alpha_d=1/\alpha_m$, independently of $T$~\cite{short}.

The effective coupling constant $g(T)$  varies with $T$ to account for
the continuously varying exponents. It can be calculated via the above
mentioned   relations  with decay  exponents  or  via fluctuations of
winding  numbers (see  Sec.~\ref{sec:windings}).  As   usual in a   CG
description, it is useful to have an  external exact result to fix the
coupling constant: here the calculations of
Refs.~\onlinecite{Kasteleyn,FisherStephenson}  give   $g(T=\infty)=1/2$.

Another insight of   the  CG picture is  about  the  relevance of  the
locking potential $V_p$  in Eq.~(\ref{eq:eff}): the identification with the
vertex   operator   of  an       electric  charge  $p$     immediately
indicates~\cite{CG}  that   it  will   become  relevant    for  $g\geq
g_c(p)=\frac{p^2}{4}$ (this is the same result as obtained within the
RG of Appendix~\ref{sec:rgsg}). In our model, $p=4$ and $g$ thus increases from
$1/2$ at $T=\infty$ to $g(T_c)=g_c(p=4)=4$ at the critical point. This
also means that locking potential terms with higher values of $p$ will
always be less relevant and can be ignored in the present model.

We finally note that an equivalent route  leading to the same CG model
is  to  consider the  low-$T$   phase.   The  system has  a  four-fold
ground-state degeneracy.   At low   $T$,  a finite  system is   mainly
composed of large domains where  all the dimers are  aligned in one of
the   four possible  columnar states  of   Fig.~\ref{fig:GS}.  One can
associate to each domain orientation a $q=4$ ``clock spin''. One could
therefore  naively  expect a  transition in  the 2D  $q=4$ state clock
model universality class.  The $q$-state  clock model is well known to
map to a Coulomb gas with (unconstrained) integer magnetic charges and
electric charges  multiples of $q$~\cite{CG}.   The  key point  here is
that due to  the absence of   monomers in our  model, there  cannot be
isolated points (sites) around which this clock-spin can make a $2\pi$
rotation passing by  the four possible  ground states.  This is nicely
illustrated in another context in Ref.~\onlinecite{Levin}. In other
words, we are considering a $q=4$ state clock model {\it with no
  topological defects}.  Magnetic charges are therefore absent in the
associated CG and we arrive at the same CG picture described above:
electric charges can only be multiples of 4 and magnetic charges are
absent.

\subsection{Transfer Matrix calculations}

TM calculations are perfectly suited to validate this CG scenario, thanks to
the conformal invariance results Eq.~(\ref{car84}) and (\ref{car86}).

\subsubsection{Fundamental exponents}
\label{sec:expo}
\begin{figure}
\begin{center}
\includegraphics[width=8cm]{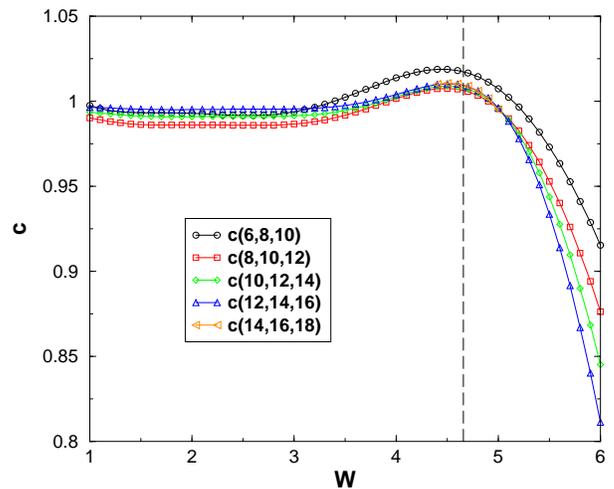}
\end{center}
\caption{(Color online) Three-point fits $c(L-4,L-2,L)$ for the central charge, as a
function of $W$. The dashed line indicates $T_c$ as found by the Monte Carlo
simulations.}
\label{fig:c3}
\end{figure}

In Fig.~\ref{fig:c3}, we show the finite-size estimates of
the central charge $c$ as a function of $W \equiv {\rm e}^{-v/T}$.
The results shown are three-point fits based on Eq.~(\ref{car86})
with a $1/L^4$ correction added. Indeed, such a non-universal
correction is predicted by conformal invariance, and is known
to greatly improve the rate of convergence. The data show a clear $c=1$ plateau for $1 \le W \le W_{\rm c}$, where
we estimate $W_{\rm c}=4.5(1)$ from the intersections of the curves.
For $W>W_{\rm c}$ the curves level off to zero, with the drop getting
sharper with increasing strip width $L$: this signals the transition to
non-critical behavior. The estimate of $W_{\rm c}$ is in perfect
agreement with the MC estimate of $T_{\rm c}$.

\begin{figure}
\begin{center}
\includegraphics[width=8cm]{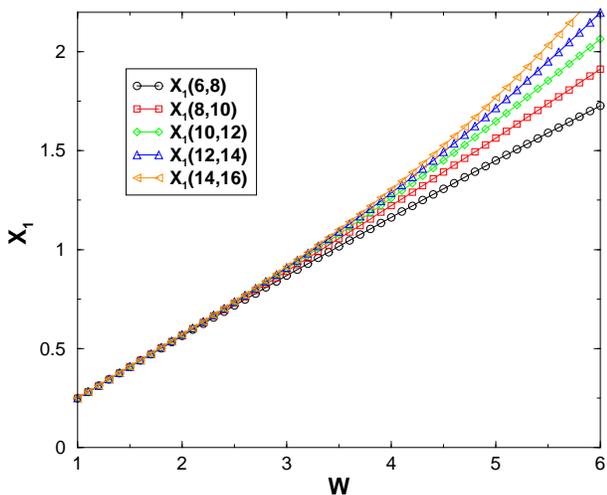}
\end{center}
\caption{(Color online) Two-point fits $X_1(L-2,L)$ for the monomer exponent, as a
function of $W$.}
\label{fig:X1}
\end{figure}

In Fig.~\ref{fig:X1} we show the finite-size estimates of the monomer
exponent $X_1 = \alpha_m/2$ as a function of $W$. The results shown
are two-point fits based on Eq.~(\ref{car84}) for $Q_0=1$ with, once
again, a $1/L^4$ correction added. The agreement with the result $X_1=1/4$
(see Section~\ref{sec:bethe.TL}) for $W=1$ is very clear. Within the
critical region $W \in [1,W_{\rm c}]$, the data can readily be extrapolated
to the thermodynamic limit $L\to\infty$, and one finds that $X_1$ is a
monotonically increasing function of $W$ that eventually takes the value
$X_1(W_{\rm c}) = 2$. So the monomer perturbation is marginal at $W_{\rm c}$
and relevant in the critical region, as predicted by the CG. For $W>W_{\rm c}$ the extrapolation
in $L$ of the numerical data does not work well, as could be expected
from the non-critical behavior predicted in this region.

\begin{figure}
\begin{center}
\includegraphics[width=8cm]{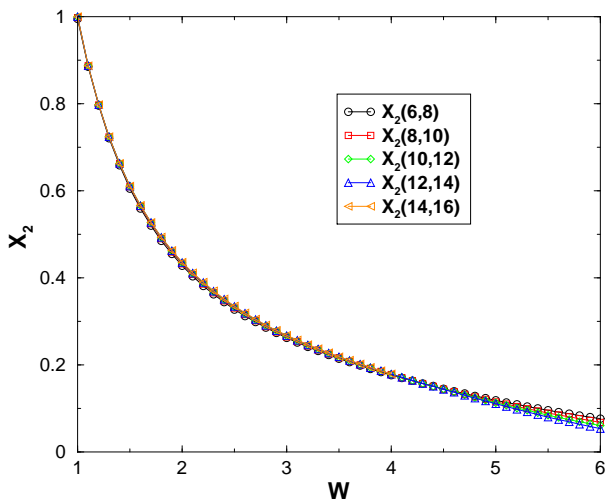}
\end{center}
\caption{(Color online) Two-point fits $X_2(L-2,L)$ for the dimer exponent, as a
function of $W$.
}
\label{fig:X2}
\end{figure}

Fig.~\ref{fig:X2} shows the finite-size estimates of the dimer
exponent $X_2 = \alpha_d/2$ as a function of $W$, based on the first
two eigenvalues in the $Q=0$ block.  There is an excellent agreement
with the result $X_2=1$ (see Section~\ref{sec:bethe.TL}) for $W=1$.  Inside the critical region
$W \in [1,W_{\rm c}]$, the data decrease monotonically, and one has to
a very good precision $X_1 X_2 = 1/4$ independently of $W$, confirming
the CG scenario involving only a single coupling constant $g$. An
estimate for the location of the critical point $W_{\rm c}$ can be
obtained from the crossings of the curves, and is consistent with the
one given above.

\begin{figure}
\begin{center}
\includegraphics[width=8cm]{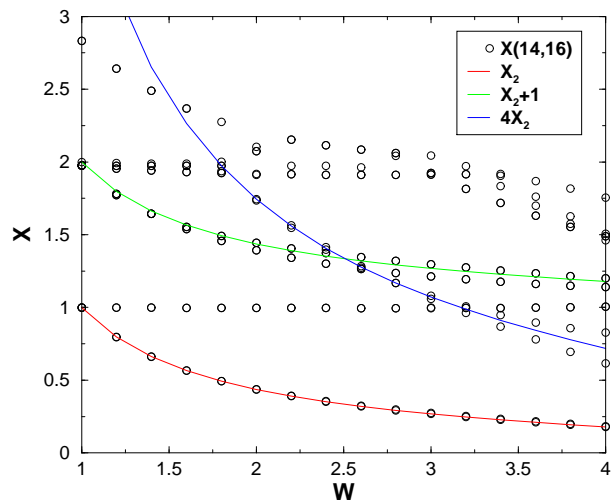}
\end{center}
\caption{(Color online) Two-point fits $X(14,16)$ for the first 15 exponents in the
  $Q=0$ block, as functions of $W$. All observed exponents are simply
  related to the fundamental dimer exponent $X_2$.}
\label{fig:Xall}
\end{figure}

\subsubsection{Higher dimer exponents}

To analyze the agreement with the CG scenario in more detail, we
show in Fig.~\ref{fig:Xall} the exponents $X$ obtained from the first
15 gaps in the $Q=0$ block. The first of these coincides with the
fundamental dimer exponent $X_2$ discussed above. This level exhibits
a two-fold degeneracy which is almost exact at finite size.
The identity operator is observed to have descendants at level 1 and 2
({\it i.e.} operators with constant $X=1$ and $X=2$), in agreement with
CFT.

More interestingly, all other exponents appear to be of the form $p^2
X_2 + q$, for integers $p \ge 2$ and $q \ge 0$. In the CG scenario,
this is accounted for by operators of electric charge $p$ times that
of the fundamental one, and their descendants. Inevitably, these
higher operators are slightly less well determined in finite size, and
the splittings of the degeneracies which would be exact in the
thermodynamic limit are somewhat larger. In the figure, we have shown
in solid line style the finite-size data for $X_2$ (corresponding to
$p=1$), $X_2+1$ (its first descendant) and $4X_2$ (the $p=2$ electric
exponent).  The agreement with the numerical data is very fine.

To summarize, the data of Fig.~\ref{fig:Xall} provide strong evidence
that the CG description of the model is correct and complete.

\subsubsection{Including monomers}
\label{sec:monomers}

The  generalization of the dimer model  to finite monomer fugacity was
first studied by TM calculations  in the grand canonical ensemble in
Ref.~\onlinecite{short}.   We  give here a   more detailed analysis. We  also  mention  a    very recent  work   by
Papanikolaou, Luijten and  Fradkin~\cite{Papa} who independently
consider the finite doping transition in this model.

The monomer exponent $X_1$ is RG relevant ({\it i.e.} $X_1 < 2$) for $1 \le
W < W_{\rm c}$. This means that the critical phase of the close-packed dimer model is
unstable toward doping with monomers. We have verified numerically
that there is no critical behavior for $W \in [1,W_{\rm c})$ and
finite monomer fugacity $\xi$. This means that the RG flow will be
toward the trivial non-critical fixed point at $\xi=\infty$.

However, $X_1$ is marginal ($X_1=2$) exactly at $W=W_{\rm c}$, and
thus one may suspect, as announced in Ref.~\onlinecite{short},
that allowing for a finite monomer density will
produce another critical line emerging from $W_{\rm c}$. To test this
suspicion, we have adapted the TM to accommodate for monomers with
Boltzmann weight $\xi$. The weight of a pair of aligned dimers is
still $W = {\rm e}^{-v/T}$ with $v=-1$.

Despite of this modification the basis states can still be described
in terms of the edge occupation numbers introduced in
Sec.~\ref{sec:bethe:int_transfer}. The main difference is in the dimer
constraint: if a vertex is occupied by a monomer, none of its four
incident edges may be occupied by a dimer. It should be noted that allowing for monomers will couple the
different blocks $T_Q$ of the TM. Accordingly, the dimension ${\rm
  dim}(T)$ of the modified TM is larger than in the pure dimer case for
which one could diagonalize sector by sector. We have therefore
restricted the study of the monomer-doped model to widths $L \le
16$.  The analytic expression for ${\rm dim}(T)$ has been given in
Eq.~(\ref{dimT}) above.

\begin{figure}
\begin{center}
\includegraphics[width=8cm]{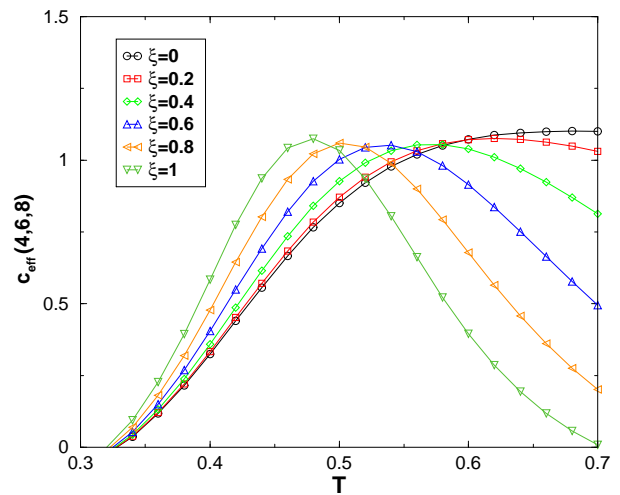}
\end{center}
\caption{(Color online) Temperature scans of the effective central charge
  $c_{\rm eff}(4,6,8)$ for various values of the monomer weight $\xi$.}
\label{fig:cmon}
\end{figure}

In Fig.~\ref{fig:cmon} we show rough scans of the effective central
charge $c_{\rm eff}$ as a function of $T$, for various values of
$\xi$ and rather small sizes $L$. We observe that for each value of
$\xi$, there exists a temperature $T(\xi)$ for which $c_{\rm eff}$
goes through a maximum.  The finite-size corrections to $T(\xi)$ are
found to be very small. We interpret the curve $T(\xi)$ as a line of fixed points
which correspond to the phase transition between high and low monomer
density.

As the maximum in $c_{\rm eff}$ becomes sharper with increasing system
size (not shown), the fixed points $T(\xi)$ are RG unstable to small variations
in the parameters $(\xi,T)$. This means that points on the low-$\xi$
side of the transition will renormalize toward vanishing monomer
density, but this phase will be non-critical (crystalline phase) since
the temperature is lower than the critical temperature in the pure
dimer model. Points on the high-$\xi$ side of the transition will
renormalize toward infinite $\xi$ as before. Note that when $\xi$ increases, $T(\xi)$ decreases. This was to be
expected, since when the dimers are diluted by more and more monomers,
it becomes harder for them to align at a given temperature.
 
\begin{table}
 \begin{tabular}{l|llllll}
 $\xi$    & 0.0    & 0.2    & 0.4    & 0.6    & 0.8    & 1.0    \\
 $T(\xi)$ & 0.6635 & 0.6125 & 0.5710 & 0.5360 & 0.5055 & 0.4795 \\ \hline
 $\xi$    & 1.4    & 1.8    & 2.6    & 4.0    & 8.0    & \\
 $T(\xi)$ & 0.4355 & 0.4015 & 0.3515 & 0.2970 & 0.2265 &
 \end{tabular}
 \caption{Phase transition temperature $T(\xi)$ in the model where monomers are allowed with a weight $\xi$. The error bar on the values of $T(\xi)$ is of the order $\pm 0.0002$.}
\label{tab:mon}
 \end{table}

\begin{figure}
\begin{center}
\includegraphics[width=8cm]{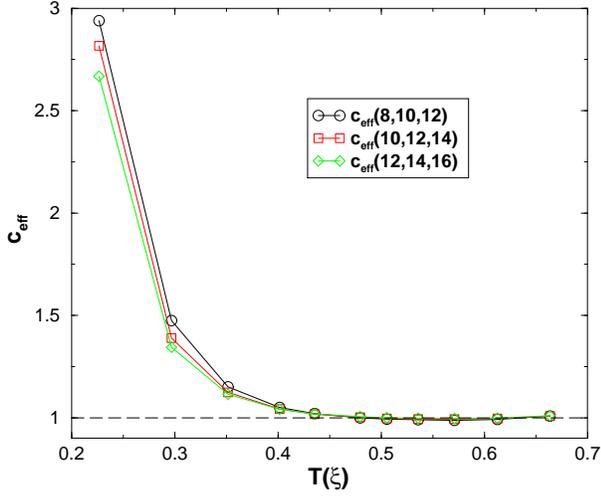}
\end{center}
\caption{(Color online) The effective central charge as a function of
  $T(\xi)$. The dashed line denotes $c_{\rm eff}=1$.}
\label{fig:c_monomer}
\end{figure}

We have determined $T(\xi)$ by taking these temperature
scans of Fig.~\ref{fig:c_monomer} to larger sizes (up to $L=16$) and carefully
studying the finite-size effects. Our final results for the phase transition
temperatures are given in Table~\ref{tab:mon}.

To determine whether the phase transition curve $T(\xi)$ corresponds to a
critical behavior, we show in Fig.~\ref{fig:c_monomer} the effective
central charge as a function of $T(\xi)$ for several different system
sizes. We observe that there exists a temperature $T_\star$ such that
$c \simeq 1$ for $T(\xi) > T_\star$. Determining $T_\star$ from this figure
is a little delicate, since even for the lowest value of $T$ shown, the
finite-size effects are such that $c_{\rm eff}$ decreases with increasing
$L$.
We can however give a first rough estimate:
\begin{equation}
\label{eq:Tstar1}
 T_\star = 0.35 \pm 0.05.
\end{equation}
We interpret the role of $T_\star$ as follows: For $T > T_\star$, the curve
$T(\xi)$ describes a line of second-order (continuous) phase transitions with
$c=1$, whereas for $T < T_\star$ the phase transition becomes first-order
(discontinuous). This means that $(\xi_\star,T_\star)$ is a {\em tricritical}
point. The tricritical nature of this point is further corroborated by an
easy domain-wall argument showing that the transition at $T=0$ is necessarily
discontinuous~\cite{short}.

\begin{figure}
\begin{center}
\includegraphics[width=8cm]{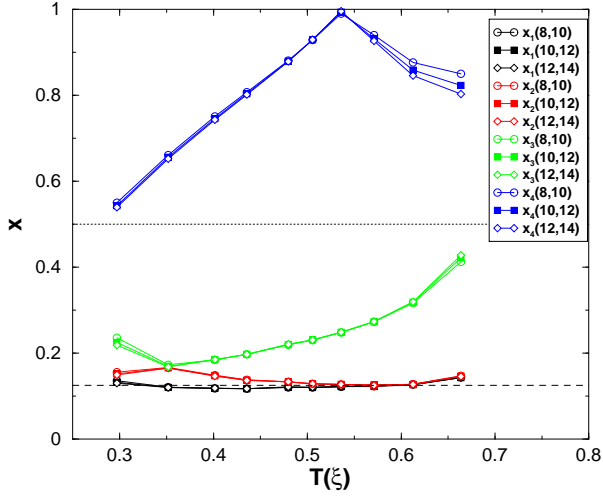}
\end{center}
\caption{(Color online) The lowest four scaling dimensions $x(L-2,L)$ along the
    critical curve $T(\xi)$. The dashed and dotted lines denote the values
    $x=1/8$ and $x=1/2$ respectively.}
\label{fig:x_monomer}
\end{figure}

To determine the universality class of the critical part of the curve
$T(\xi)$, we show in Fig.~\ref{fig:x_monomer} estimates for the lowest four
scaling dimensions (denoted $x_1 \le x_2 \le x_3 \le x_4$) as functions of
$T(\xi)$. A conspicuous feature is that $x_1 \simeq 1/8$ is almost independent of
$T(\xi)$, and that $x_2$ is almost degenerate with $x_1$. The existence of a
constant exponent $x=1/8$ (first noted in Ref.~\onlinecite{short}) and other interaction-dependent exponents, is
characteristic of the two-dimensional Ashkin-Teller \cite{AT} (AT) model.
We therefore conjecture that the critical curve $T(\xi)$ is {\em in
the universality class of the AT model}.

To test this conjecture and make it more precise, we recall some facts about
the isotropic, selfdual AT model, following section 12.9 of Baxter's book
\cite{Baxter_book}. The AT model is originally defined in terms of two Ising
models, each with coupling constant $K$, that interact via a four-spin
(energy-energy) coupling $K_4$. Defining $\omega_1 = \exp(-K_4)$ and
$\omega_3=\exp(K_4-2K)$, one finds, after a long series of
transformations, that the model is equivalent to a six-vertex model with
parameters $a=b=\sqrt{2} \, \omega_1$ and $c=\sqrt{2} \, (\omega_1+\omega_3)$.
The anisotropy parameter $\Delta$ of the corresponding XXZ spin chain reads
\begin{equation}
 \Delta = 1 - \frac12 \left( 1+\frac{\omega_3}{\omega_1} \right)^2
        = 1 - 2 \cos^2\left( \frac{\pi y}{4} \right) \,,
\end{equation}
where the second equality parametrizes only the critical regime
$|\Delta| \le 1$, through the parameter $y \in [0,2]$. We note
that $y=0$ corresponds to a KT transition. The original couplings, $K$
and $K_4$, are only real and finite for $y \in [0,4/3)$, but from the point of
view of the six-vertex model nothing special happens at $y=4/3$. The critical
exponents of the AT model differ subtly from those of the equivalent
six-vertex model. There are two magnetic-type exponents, which in our notation
read $x_H^{(1)} = 1/8$ and $x_H^{(2)} = \frac{1}{8-4y}$, and a
temperature-like exponent $x_t=\frac{1}{2-y}$.

We now claim that the critical line $T(\xi)$ corresponds to the AT model with
$y \in [0,3/2]$. Moreover, we identify $x_1 = x_2$ of
Fig.~\ref{fig:x_monomer} with $x_H^{(1)}$ in the AT model, $x_3$ with
$x_H^{(2)}$, and $x_4$ with $x_t$ (the latter identification is only on the
low-$T$ side of the level crossing visible in the figure; for higher $T$ we
identify $x_t$ with the analytic continuation of $x_4$). The parameter $y$
is an (unknown) increasing function of $T$, taking the values $y=0$ at
$T=T_\star$ and $y=3/2$ at $T=T_{\rm c}$.

Using Fig.~\ref{fig:x_monomer}, we can now identify $T_\star$ either from
$x_3 = 1/8$ or from $x_4 = 1/2$. These two determinations are consistent,
but the latter leads to the best final estimate of the tricritical point:
\begin{equation}
 T_\star = 0.29 \pm 0.02 \,,
\end{equation}
more precise than the first estimates of Eq.~\ref{eq:Tstar1} and of Ref.~\onlinecite{short}. 
For $y=0$, the original couplings of the AT model satisfy $K=K_4$, meaning
that the two constituent Ising models couple strongly so as to give a single
four-state Potts model. We conclude that the transition at $T_\star$ is in the universality class
of the critical ferromagnetic four-state Potts model. Note that this
identification is consistent both with the tricritical nature of the
transition, and with the existence of a RG marginal direction in the
Potts model~\cite{NBRS}.

Meanwhile, for $y=3/2$ we have $x_H^{(1)} = 1/8$ and $x_H^{(2)} = 1/2$. This
is in nice agreement with the values of the first two electric-type
exponents as obtained from the Coulomb gas of the pure dimer model
at $T=T_{\rm c}$. Moreover, $x_t=2$ becomes marginal and can be identified
with the monomer operator, which is responsible for the transition.

As a last check of the AT identification, consider the point $y=1$, where
$K_4=0$ and $K=\frac12\ln(1+\sqrt{2})$, so that the model decouples into two
non-interacting critical Ising models. Here one has $x_H^{(2)}=2 x_H^{(1)}=1/4$
and $x_t = 1$. One can verify from Fig.~\ref{fig:x_monomer} that this indeed
happens at the same temperature, $T_{\rm Ising} \simeq 0.54$.

The analysis presented in this section is in agreement with the results of
Ref.~\cite{Papa}.

\subsection{Winding number fluctuations and Coulomb gas constant}
\label{sec:windings}

In this section, we derive an analytical relationship between the
coupling constant $g$ of the CG in the high-$T$ region and the
fluctuations of dimer winding numbers (to be defined below). As
winding number fluctuations are easily accessible in MC simulations,
this allows for an independent calculation of the CG coupling constant
that can be compared with TM calculations.

Orienting the dimers from the odd sublattice toward the even sublattice
defines a fictitious ``magnetic field'' $\vec B$ on each bond~\cite{Huse}. Because each
site has exactly one dimer, the lattice divergence of this magnetic field
is ${\rm div} \vec B=1$ (resp. -1) on the even (resp. odd) sublattice. As a
consequence (Gauss law), the flux of $\vec B$ through a contractible loop
(with as many sites from each sublattice) is zero in any dimer
covering~\cite{monomer.note}. However, the fluxes $W_x$ and $W_y$ through the two (non-contractible)
cycles winding around the torus are two non-trivial integers
In the following, we denote $P(W_x,W_y)$ the probability  to observe a
dimer configuration with winding numbers $W_x,\ W_y$.

 In the continuum limit, this probability can be obtained in the
high-$T$ regime where the dimer configurations are
coarse-grained into a free field~\cite{free}. In the height representation of
Sec.~\ref{sec:height}, the field $h$ is defined by the
integral of its derivative. The winding number across a cycle is
given by this integral along the cycle. The probability
$P(W_x,W_y)$ is the ratio of the partition function restricted to
fields having winding numbers equal to $W_x,W_y$ across the two
cycles of the torus: $Z_{W_x,W_y}$, divided by the total partition
function $\sum_{W_x,W_y\in {\mathbb Z}}Z_{W_x,W_y}$.

This ratio can be evaluated as follows: one separates the height
field into a classical and a fluctuating part $h=h_{\rm
cl}+\delta$ where $h_{\rm cl}$ is the solution of the equations of
motion (a harmonic function) carrying the two winding numbers
$W_x,W_y$, and $\delta$ is a fluctuating field with no
discontinuity. $h_{\rm cl}$ being a solution of the
equations of motion, the crossed term disappears from the action
and the free part of the action Eq.~(\ref{eq:eff}) is the sum of a classical part and a
fluctuating part which does not depend on the winding numbers. As
a result, the partition function $Z_{W_x,W_y}$ factorizes into a
classical part $Z^{\rm cl}_{W_x,W_y} $ and a fluctuating part
$Z'$, which being independent of the winding number, factorizes out of the
probability $P(W_x,W_y)$.

On a square torus of size $L_x,L_y$, the classical height
configurations are given by:
\begin{eqnarray}
h(x,y)=x{W_x\over L_x}+y{W_y\over L_y},
\end{eqnarray}
and the probability is obtained by substituting this field in the
expression of the Boltzmann weight:
\begin{eqnarray}
 P(W_x,W_y)={  e^{-\pi g (({W_x\over L_x})^2+({W_y\over L_y})^2)L_x L_y}
 \over \sum_{n,m\in {\mathbb Z}}
	e^{-\pi g(({n\over L_x})^2+({m\over L_y})^2)L_x L_y}}.
  \label{proba}
\end{eqnarray}
The quadratic fluctuation of the winding number across the
direction $x$  is thus given by:
\begin{eqnarray}
\langle W_x^2\rangle={\sum_{n\in {\mathbb Z}}  n^2 e^{-\pi g n^2 (L_y/L_x)}
 \over \sum_{n\in {\mathbb Z}} e^{-\pi g n^2 (L_y/L_x)}}.
  \label{eq:wind}
\end{eqnarray}

In the high-$T$ phase of the dimer model, {\it short-distance}
properties  are of course {\it not}  described by  a free height field.
However, a  local dimer move cannot change  the winding  number and we
expect that the       precise (non-Gaussian) nature  of   the    local
fluctuations   will not affect  winding   number observables. $\langle
W_x^2\rangle$ is related to (fluctuations of)  the global slope of the
height profile. It can be expressed using the Fourier modes $|k|\ll 1$
of  the height field only~\cite{Henley} and  these are precisely those
described by the Gaussian action.

\begin{figure}
\begin{center}
\includegraphics[width=8cm]{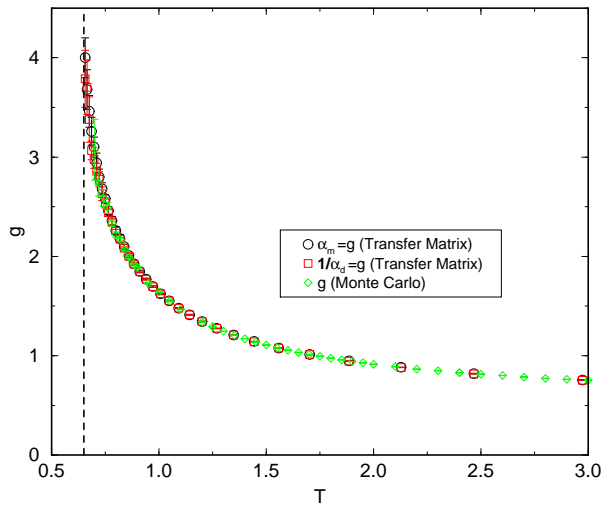}
\caption{(Color online) Coulomb gas coupling constant $g$ as a function of
  temperature $T$ obtained via TM (through exponents $X_1$ and $X_2$ - see Sec.~\ref{sec:expo}) and MC calculations (thanks
  to Eq.~\ref{eq:wind}).}
\label{fig:g}
\end{center}
\end{figure}

In the non interacting case ($T=\infty$) where $g=1/2$, Eq.~(\ref{eq:wind}) has recently been derived using the
Pfaffian expression of the partition function by C.~Boutillier and
B.~de Tili\`ere~\cite{Boutillier}. For other values of $T$, it allows to calculate through MC simulations the temperature
dependence of the coupling constant $g$, which can be compared with the
one obtained from dimer and monomer exponents calculated via the
TM (see Sec.~\ref{sec:expo}). The temperature dependence of $g(T)$ obtained by the three
methods match perfectly as can be seen in Fig.~\ref{fig:g}, where the three curves basically overlap.  The value of $g$ at
$T_c$ (denoted by a dashed line in the figure) is $g_c=g(T_c)=4.0(2)$, in agreement with the CG prediction $g_c=4$.
The CG analysis and the free field calculation Eq.~(\ref{eq:wind}) are therefore
entirely validated.

\section{Discussion}
\label{sec:discuss}

\subsection{Connection to other classical models}

With the interpretation of Sec.~\ref{sec:CG}, the interacting classical
dimer model is naturally related to other scalar-field models displaying a CG
behavior~\cite{CG}. Even if it is not probably exactly solvable, its
definition puts it among the most simple ones, along with the XY
or 6-vertex models. Moreover, it can be quite naturally extended ({\it
  e.g}. by changing the sign of interactions, going to other geometries), which allows to test various features of the CG. 

The model Eq.~(\ref{eq:model}) also admits a height description
(Sec.~\ref{sec:height}), which relates it to a variety of spin and
Solid-On-Solid models (see Ref.~\onlinecite{Henley} and references therein)
which display the same physical behavior. We believe this interacting dimer
model offers several advantages over these models: (i) its definition is
quite simple and natural, (ii) it admits two exacts points (at zero and
infinite temperature), (iii) {\it very efficient} numerical simulations are
possible (thanks to the directed-loop algorithm~\cite{Sandvik} for the MC
simulations and to the relative small size of the configuration
space for the TM calculations).

Concerning numerical simulations of height-like models, we find that the
height description is not strictly necessary to obtain reliable numerical
results (as opposed to Ref.~\onlinecite{Henley}c). Henley and
coworkers~\cite{Henley} find the height representation particularly useful
because it allows for example to calculate the stiffness (or CG) constant
$g$ via the long-wave lengths fluctuations of the Fourier transform of the
height field. We have proposed here an alternative method to obtain $g$ via
the fluctuations of the winding number (see Sec.~\ref{sec:windings}), which
has the advantage of not involving any fitting procedure.

Concerning the physics of classical dimers, the analysis presented here can
probably be extended to the recent results of Sandvik and Moessner~\cite{Sandvik} on
non-interacting models with longer-range dimers. For models with
dimers allowed between next-nearest neighboring sites, the bipartite nature
of the lattice is lost and the (dislocation-free) height mapping no longer valid: correlations
become exponential. For models with a fraction of dimers allowed between
fourth-nearest-neighbors, both bipartiteness and height descriptions are
present: the system stays critical. There, Ref.~\onlinecite{Sandvik} finds
that the monomer-monomer decay exponent continuously varies (from
$\alpha_m=1/2$ to $\alpha_m=1/9$) with the fraction (fugacity $w_4$) of
longer-range dimers: this is naturally explained in our framework by
considering the variation of the CG constant $g$ with $w_4$ - which in turns
controls the monomer-monomer decay exponent. It it also found that the
dimer-dimer correlation keeps its $1/r^2$ behavior. This last finding was
accounted for in Ref.~\onlinecite{Sandvik} by considering the ``dipolar''
terms in the dimer-dimer correlation functions. The
corresponding part of the dimer-dimer correlation function continues to vary
as $r^{-2}$ whereas the vertex contribution~\cite{Fradkin} scales as
$r^{-1/g(w_4)}$. The long-distance behavior observed in the numerics is
therefore dominated by the largest exponent between both; here it is
$2$. This is reminiscent of spin-spin correlation functions of the XXZ
quantum spin chain~\cite{Luther} where the  uniform part
of the $\langle S^z(0)S^z(r)\rangle $ correlation function decays as $1/r^2$
whereas the staggered part decays as $1/r^{f(\Delta)}$ where $f$ is a
continuous function of the anisotropy parameter $\Delta$.

Finally, we remind that the interacting dimer model Eq.~(\ref{eq:model}) was
introduced in the context of liquid crystal physics. By means
of a low-$T$ expansion, Poland and Swaminathan estimated in
Ref.~\onlinecite{old}b the transition temperature to be $T_c=0.61(1)$, quite
close to the actual result. However, they incorrectly stated
at that time that the transition was Ising-like (second order) and did not
realize the critical nature of the high-$T$ phase.

\subsection{Implications for finite temperature properties of the Quantum Dimer Model on the square lattice}

The classical model naturally offers informations about the finite-$T$
properties of QDM~\cite{rk}, where quantum effects are not dominant. In
particular, the critical phase found in our model should be present in all
the high-$T$ phase diagram of the QDM: indeed, the rough nature of this phase
was shown to be intimately tied to the dimer hard-core constraint and not to
the details of fluctuations (thermal or quantum). We note in passing that,
as in the classical case, doping the QDM with monomers (static or mobile)
will immediately destroy this critical phase~\cite{Krauth}. We also think
that the finite-$T$ melting of the columnar phase will proceed via a
KT transition as described here everywhere in the  phase
diagram of the QDM (presumably, this is also the case for the melting of the
plaquette phase). We note that these findings are in contrast to the
finite-$T$ phase diagram of the QDM proposed by Leung and
coworkers~\cite{Leung}, who speculated that the columnar crystal would melt
in two steps (with an intermediate plaquette phase), even in the classical
limit.

The $T=0$ phase diagram of the QDM is usually accepted as
this~\cite{Leung,Sylju05a,Sylju05b} (with the standard notation of $t$ for the
kinetic energy gained by flipping a plaquette): for large negative $v/t$,
the system is in a columnar phase. Increasing $v/t$, the QDM experiences a
quantum phase transition to a plaquette phase at a quite uncertain value of
$v/t$ (see discussion below). The plaquette phase ends up at the
Rokhsar-Kivelson (RK) multi-critical point $v/t=1$, where the correlation
functions display algebraic quasi-long range order~\cite{rk}. For $v/t>1$,
the system is in the staggered phase. 

\begin{figure}
\begin{center}
\includegraphics[width=8cm]{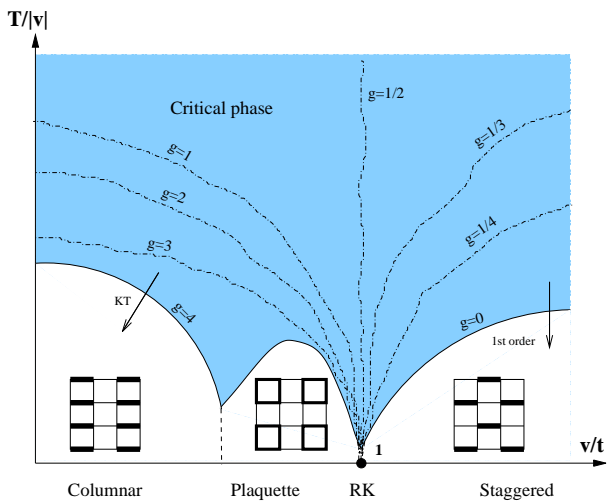}
\caption{(Color online) Schematic finite-temperature phase diagram of the Quantum Dimer Model on the square lattice. The different crystals (columnar, plaquette and staggered) all melt into a high temperature critical phase (denoted in blue) with algebraic correlations characterized by a Coulomb gas constant $g$. See text for details. }
\label{fig:QDM}
\end{center}
\end{figure}

We are now armed to connect this phase diagram to our finite-$T$ results for
$t=0$ and draw our proposed $(v/t,T/|v|)$ schematic phase diagram (see
Fig.~\ref{fig:QDM}). We believe the critical phase present in the
whole high-$T$ region extends down to low temperatures at the RK point. This
critical phase can be parametrized by the Coulomb gas coupling constant $g$,
and we draw lines of iso-$g$ in this region. The line $g=4$ denotes the
boundary between the high-$T$ phase and the columnar phase (and presumably to
the plaquette phase as well), corresponding to the expected KT
transition. As the transition from the plaquette to the columnar phase is
expected on general grounds to be first order at $T=0$, we expect the
transition to subsist at finite temperature. On the right hand side of the phase diagram, the rough phase
will give rise to the staggered phase at low temperatures via a vanishing of
the CG constant $g\rightarrow 0$. First results indicate that the transition
is here first order~\cite{Castelnovo,note.v1}. Finally, we conjecture that all iso-$g$
lines meet at the RK point $(T=0,v/t=1)$ (see Fig.~\ref{fig:QDM}),
consistent with the multi-critical nature of the RK
point~\cite{Fradkin}. Even though numerically difficult (see discussion
below), it would be very interesting to test these predictions, especially
in the vicinity of the RK point.

We now make a remark on the existence of the plaquette phase. We first note
that our classical model is amenable to large-scale simulations, which is
not the case of the QDM. Numerical simulations of the QDM are indeed very
difficult: exact diagonalizations~\cite{Sachdev,Leung} are limited to small
sizes (at most $10 \times 10$ lattices) and even if the sign problem is absent in
the QDM, Quantum Monte Carlo (QMC) calculations are notoriously hard. Some
progresses have however recently been made in the Green Function QMC
formulation~\cite{Sylju05a,Sylju05b,Ralko}. These difficulties could have
important consequences for the plaquette phase: indeed, we saw that
plaquette correlations were important in our model and altered finite-size
behavior, even on systems as large as $160 \times 160$. This is an indication
that the plaquette phase in the QDM could as well just disappear in the
thermodynamic limit. This is not so
unlikely noticing that the critical point separating the columnar and
plaquette phase was first reported~\cite{Leung} to be at $v/t \sim -0.2$,
and then presumably around $v/t =0.60(5)$ when larger samples were
available~\cite{Sylju05a,Sylju05b}: this indicates that the
plaquette phase extent actually shrinks when increasing system size.

\section{Conclusion}
\label{sec:conc}

In this work, we studied a model of interacting close-packed dimers on
the square lattice, where the interaction tends to align neighboring
dimers. The ground-state constitutes of dimers aligned in column and
is four-fold degenerate. The $T=\infty$ point is the dimer coverings
enumerating problem, and displays algebraic correlations. With the
help of MC and TM simulations, we could show that the low-$T$
crystalline columnar phase leaves place to a high-$T$ critical phase,
with floating decay exponents of the correlation functions. The
transition takes place at $T_c=0.65(1)$ and is of the KT type. Via a
height description of the dimer configurations, we determined the
effective continuum theory with a competition between a stiffness term
which favors rough height profiles (critical phase for the dimers),
and a locking potential favoring heights to be locked on four
particular values (forcing the dimers to align in one of the four
columnar ground states). This description is equivalent to a CG of
electric charges, and the locking potential is in this picture
associated with the vertex operator of $e=4$ electric charges. This
allows us to interpret all the numerical results obtained in the
high-$T$ phase in terms of the CG coupling constant $g$, which we
calculate numerically with high precision. The transition is
understood in the CG language as a proliferation of $e=4$ electric
charges. This model is probably one of the simplest and versions of a
CG with floating exponents. Finally, we have established that doping
the dimer model with monomers produces another critical line emanating
from the KT point. There is strong numerical evidence that this line
is in the universality class of the Ashkin-Teller model. The AT line
terminates in a tricritical point at finite $T_\star = 0.29(2)$, in the
four-state Potts universality class. Below $T_\star$ the
transition goes first-order. Besides their relevance to other
classical models, the results of this work also have important
implications for QDM on bipartite lattices (see
Sec.~\ref{sec:discuss}). In particular, it indicates that their
high-$T$ phase is also critical, and we conjecture that this
criticality extends down to the RK point.

\begin{acknowledgments}
We thank C.~Boutillier, E.~Fradkin, C.~L.~Henley, W.~Krauth, R.~Moessner,
D.~Poilblanc, P.~Pujol, A.~Ralko and O.~Sylju\aa sen for fruitful discussions and gratefully acknowledge
Fr\'ed\'eric Mila and Matthias Troyer for their participation at an early
stage of this work. F.A. is supported by the French ANR program. The MC simulations were performed on the
Gallega cluster at SPhT using the ALPS libraries~\cite{ALPS}. 

\end{acknowledgments}
\appendix
\section{Perturbative renormalization group for the 2D sine-Gordon model}
\label{sec:rgsg}

We follow here the presentation given by Levitov\cite{levitov}.  
We write the energy of the sine-Gordon model as:

\begin{equation}
	E=\int d^Dr \left[
		\frac{1}{2}\left(\nabla\Phi\right)^2+\lambda\cos(\beta\Phi)
	\right].
	\label{eq:EPhi}
\end{equation}

We introduce two cut-offs $\Lambda$ and $\Lambda'$ in momentum space
so that the field $\Phi$ splits into fast and slow components:
\begin{eqnarray}
	\Phi(k)&=&\Phi(k)^>+\Phi(k)^<
\end{eqnarray}
where $\Phi(k)^>$ (fast)   is non-zero only if  $\Lambda'\leq  |k|\leq
\Lambda$ and  $\Phi(k)^<$ (slow)   is  non-zero only  if $0  \leq |k|<
\Lambda'$.

As usual we integrate over the fast component $\Phi^>$
to obtain a renormalized energy for the slow degrees of freedom:
\begin{eqnarray}
	e^{-E^{\Lambda'}(\Phi^<)}&=&
	\int \mathcal{D}\left[\Phi^>\right]e^{-E(\Phi^<+\Phi^>)}.
\end{eqnarray}
It is simple to see that the elastic part of the energy does not couple
the            slow             and            fast         components
\begin{equation}
	\left(\nabla\Phi\right)^2
	=\left(\nabla\Phi^>\right)^2+\left(\nabla\Phi^<\right)^2
\end{equation}
so that we have:
\begin{eqnarray}
	e^{-E^{\Lambda'}(\Phi^<)}&=&e^{-\frac{1}{2} \int d^Dr (\nabla\Phi^<)^2}
	\nonumber \\
	&&\times\int \mathcal{D}\left[\Phi^>\right]
	\exp\left(-\frac{1}{2}\int d^Dr (\nabla\Phi^>)^2 \right. \nonumber \\
	&&\left.
	+\lambda\int d^Dr\cos(\beta(\Phi^>+\Phi^<))\right).
\end{eqnarray}
In the limit where the cosine term can be treated perturbatively
we have:
\begin{eqnarray}
	&&e^{-E^{\Lambda'}(\Phi^<)+\frac{1}{2}\int d^Dr (\nabla\Phi^<)^2}\simeq
	\int \mathcal{D}\left[\Phi^>\right]e^{-\frac{1}{2}\int d^Dr (\nabla\Phi^>)^2 }
	\nonumber \\
	&&\times
	\prod_r\left(1+\lambda\cos(\beta(\Phi^>(r)+\Phi^<(r)))\right) \nonumber \\
	&\sim&
	\prod_r\left(1+\lambda\left<\cos(\beta(\Phi^>(r)+\Phi^<(r)))\right>_{\Phi^>}\right) \nonumber \\
	&\sim&
	\exp\left(\lambda\left<\int d^Dr\cos(\beta(\Phi^>(r)+\Phi^<(r)))\right>_{\Phi^>}\right),
\end{eqnarray}
and finally:
\begin{eqnarray}
	E^{\Lambda'}(\Phi^<) &\simeq&\frac{1}{2}\int d^Dr (\nabla\Phi^<)^2
	\nonumber \\ 
	&&+\lambda\left<\int d^Dr\cos(\beta(\Phi^>(r)+\Phi^<(r)))\right>_{\Phi^>}
	\label{eq:E}
\end{eqnarray}
where $\left<\cdots\right>_{\Phi^>}$   is  an expectation   value with
respect   to    the   (Gaussian)  weight  $e^{-\frac{1}{2}\int    d^Dr
(\nabla\Phi^>)^2}$.  This expectation  value  can be  computed in  the
following way:
\begin{eqnarray}
\left<\cos(\beta(\Phi^>(r)+\Phi^<(r)))\right>_{\Phi^>}=\nonumber \\
	\frac{1}{2}e^{i\beta\Phi^<(r)}\left<e^{i\beta\Phi^>(r)}\right>_{\Phi^>} +{\rm H.c.} \label{eq:cos}
\end{eqnarray}
and
\begin{eqnarray}
\left<e^{i\beta\Phi^>(r)}\right>_{\Phi^>}\sim
	\int \mathcal{D}\left[\Phi^>\right]
	\exp\left(\int_{\Lambda'<|k|<\Lambda}\frac{d^Dk}{(2\pi)^D}
	\left\{
	\right.\right.\nonumber \\\left.\left.
	-\frac{k^2}{2}\Phi^>_k \Phi^>_{-k} +i\beta\Phi^>_ke^{ikr} \right\}\right).
\end{eqnarray}
The last expression is a product of Gaussian integrals. The result is:
\begin{eqnarray}
\left<e^{i\beta\Phi^>(r)}\right>_{\Phi^>}\sim
	\exp\left(-\frac{\beta^2}{2}\int_{\Lambda'<|k|<\Lambda}\frac{d^Dk}{(2\pi)^D}\frac{1}{k^2}\right).
\end{eqnarray}
Combining this result with Eqs.~(\ref{eq:E}) and (\ref{eq:cos}) we get:
\begin{eqnarray}
E^{\Lambda'}(\Phi^<)&\simeq&\frac{1}{2}\int d^Dr (\nabla\Phi^<)^2
	\nonumber \\ 
	&&
	+\lambda^*\int d^Dr\cos(\beta(\Phi^<(r)))
\end{eqnarray}
with
\begin{equation}
	\lambda^*=\lambda 
	\exp\left(-\frac{\beta^2}{2}\int_{\Lambda'<|k|<\Lambda}\frac{d^Dk}{(2\pi)^D}\frac{1}{k^2}\right).
\end{equation}

If $D=2$ we have:
\begin{equation}
	\int_{\Lambda'<|k|<\Lambda}\frac{d^2k}{(2\pi)^2}\frac{1}{k^2}=\frac{1}{2\pi}\log\frac{\Lambda}{\Lambda'}
\end{equation}
and thus
\begin{equation}
\lambda^*=\lambda \left(\frac{\Lambda'}{\Lambda}\right)^{\beta^2/(4\pi)}.
\end{equation}
The elastic term is  of the order  of $\Lambda^2$ for $\Phi^>$  and of
order  $\Lambda'^2$ for $\Phi^<$. By    integrating out $\Phi^>$,  the
importance   of  the  elastic term has    been   reduced by   a factor
$\left(\frac{\Lambda'}{\Lambda}\right)^2$. The strength of the  cosine
potential relative to the elastic terms has thus been multiplied by
\begin{equation}
\left(\frac{\Lambda'}{\Lambda}\right)^{\beta^2/(4\pi)-2}
\end{equation}
when going from $E$ to $E^{\Lambda'}$.  Thus,  we find that the cosine
term is {\em irrelevant} if $\beta^2>8\pi$. On the other hand,
the systems goes into a locked phase when $\beta^2<8\pi$.

To    make  contact with   Eq.~(\ref{eq:eff}),  we   take the  following
normalization:
\begin{equation}
	E=\int d^Dr \left[
		\pi g\left(\nabla h\right)^2+V_p\cos(2\pi p\;h )
	\right]
\end{equation}
and we  find that  it   is    equivalent to  Eq.~(\ref{eq:EPhi})    provided
$\beta=p\sqrt{\frac{2\pi}{g}}$.  With  this  normalization, the cosine
term is relevant  when $g>p^2/4$.  For  $p=4$ (relevant for the square
lattice dimer  model), we obtain that  the cosine is relevant when the
stiffness $g$ exceeds $g_c=4$.

\section{Mapping of the one-component Coulomb gas to the sine-Gordon model}
\label{sec:coulomb2sg}

This derivation is  a slightly more detailed  version of  the argument
presented  in chapter 4 of Polyakov's book~\cite{polyakov}. We  start
from a system of charges interacting with a Coulomb potential $V(r)$:
\begin{equation}
	Z=\sum_N\frac{\xi^N}{N!}\sum_{\{q_a\}}\sum_{\{r_a\}}
		\exp\left(-\sum_{a\ne b}q_aq_b V(r_a-r_b)\right)
\end{equation}
where $N$  is the number   of  charges, $\xi$ their  fugacity,   $r_a$
($a=1\cdots N$) their  coordinates on a $D$-dimensional  cubic lattice
(with unit lattice spacing)  and $q_a\in\mathbb{Z}$ their charges.

We write the interaction in momentum space:
\begin{eqnarray}
	\sum_{a\ne b}q_aq_b V(r_a-r_b) &=&\sum_{r,r'} q(r)q(r')V(r-r') \\
				       &=&\int \frac{d^Dk}{(2\pi)^D} q(k)q(-k)V(k)
\end{eqnarray}
where
\begin{equation}
	q(r)=\sum_{a=1}^N q^a \delta(r-r_a)
\end{equation}
is the charge density at $r$ and the Coulomb potential is
\begin{equation}
	V(k)=\frac{K}{4k^2}.
\end{equation}
We decouple the charges (Hubbard-Stratonovich)  by introducing a  real
scalar field $\chi(r)$:
\begin{eqnarray}
	Z&=&\sum_N\frac{\xi^N}{N!}\sum_{\{q_a\}}\sum_{\{r_a\}}\int{\mathcal{D}[\chi(r)]}
	\exp\left(-\frac{1}{K}\sum_r (\nabla \chi(r))^2
	\right.\nonumber\\
		&&\left. +i\sum_i q(r_i)\chi(r_i) \right).
\end{eqnarray}
Then we   will  perform the summation over    the particle  degrees of
freedom: their  number  $N$, their     charges  $q^a$ and    locations
$r_a$.   This  is just a sum over the charge
density $q(r)$:
\begin{eqnarray}
	\sum_{N,\{q_a\},\{r_a\}}\frac{1}{N!}= \sum_{\{q(r)\}}.
\end{eqnarray}
In    the  limit of   small   fugacity $\xi\ll1$, configurations with
$|q(r)|>1$   are exponentially   suppressed,  and   we  can keep    only
$q(r)\in\{-1,0,1\}$.   The number    of  particles is  thus  $N=\sum_r
q(r)^2$. We have to evaluate:
\begin{eqnarray}
\sum_{q(r)\in\{-1,0,1\}} \xi^{\sum_r q(r)^2} e^{i\sum_r \chi(r)q(r)}
		\label{eq:q01} \\
	       =\prod_r\left(
			\sum_{q=-1}^1 	\xi^{q^2} e^{i\chi(r)q}
			\right) \\
	       =\prod_r\left(
			1+2\xi\cos \chi(r)
			\right) \\
	       \simeq\exp\left(2\xi\sum_r\cos \chi(r)\right).
\end{eqnarray}
In the limit $\xi\ll1$, the partition function is that of the sine-Gordon model:
\begin{equation}
	Z\simeq \int{\mathcal{D}[\chi(r)]}
	\exp\left( -\frac{1}{K}\sum_r (\nabla \chi(r))^2 +2\xi\sum_r\cos \chi(r)\right).
	\label{eq:sg}
\end{equation}


\begin{references}
\bibitem{note.corners} The  four corners in fact have the same height.
\bibitem{note.g} $g$ is related to the stiffness constant $K$ used in Refs.~\onlinecite{Henley} by $g=8K/\pi$.
\bibitem{note.exponent} By exponent, we denote the actual exponent of a
  charge-charge correlation function, not the scaling dimension $X(e,m)$ of
  the electromagnetic operator. We have the simple relation $\alpha (e,m)=  2 X (e,m)$.
\bibitem{monomer.note} If monomers (magnetic charges) were allowed, the flux
would count the number of monomers on the even sublattice (magnetic charge
+1) minus the number of monomers on the odd sublattice (charge -1).
\bibitem{note.v1} The MC simulations are difficult in this case, and for
  $v/t\rightarrow \infty$ a transition happening only at $T=0$ cannot be
  excluded. This would reduce the occupation of the phase diagram by the staggered phase to a simple $T=0$ line.
\end{references}
\end{document}